\RequirePackage{fix-cm}
\documentclass[smallextended]{svjour3}       
\smartqed  
\usepackage{graphicx}
 \journalname{}
\begin{document}

\title{Semi-inverse method in nonlinear mechanics: application to couple shell- and beam-type oscillations of single-walled carbon nanotubes.
}

\titlerunning{Strong mode coupling in vibrations of  single-walled carbon nanotubes}        

\author{V.V. Smirnov        \and
        L.I. Manevitch 
}


\institute{V.V. Smirnov  \and
           L.I. Manevitch \at
              Institute of Chemical Physics, RAS,
4 Kosygin str., 119991 Moscow, Russia \\
              Tel.: +7 495-939-7515\\
             \email{vvsr@polymer.chph.ras.ru}           
}


\maketitle

\begin{abstract}
We demonstrate the application of the efficient semi-inverse asymptotic method to resonant interaction of the nonlinear normal modes belonging to different  branches of the CNT vibration spectrum.
Under condition of the 1:1 resonance of the beam and circumferential flexure modes we obtain the dynamical equations, the solutions of which describe the coupled stationary states.
The latter are characterized by the non-uniform distribution of the energy along the circumferential coordinate.
The non-stationary solutions for obtained equations correspond to the slow change of the energy distribution.
It is shown that adequate description of considered resonance processes can be achieved in the domain variables.
They are the linear combinations of the shell- and beam-type normal modes.
Using such variables we have analyzed not only nonlinear normal modes but also the limiting phase trajectories describing the strongly non-stationary dynamics.
The evolution of the considered resonance processes  with the oscillation amplitude growth is analyzed by the phase portrait method and verified by the numerical integration of the respective dynamical equations.

\keywords{Carbon nanotubes \and Nonlinear oscillations\and Mode coupling \and Energy exchange \and Energy localization \and Semi-inverse method}
\end{abstract}

\section{Introduction}\label{Int}
\begin{figure}\centering{
\includegraphics[width=50mm]{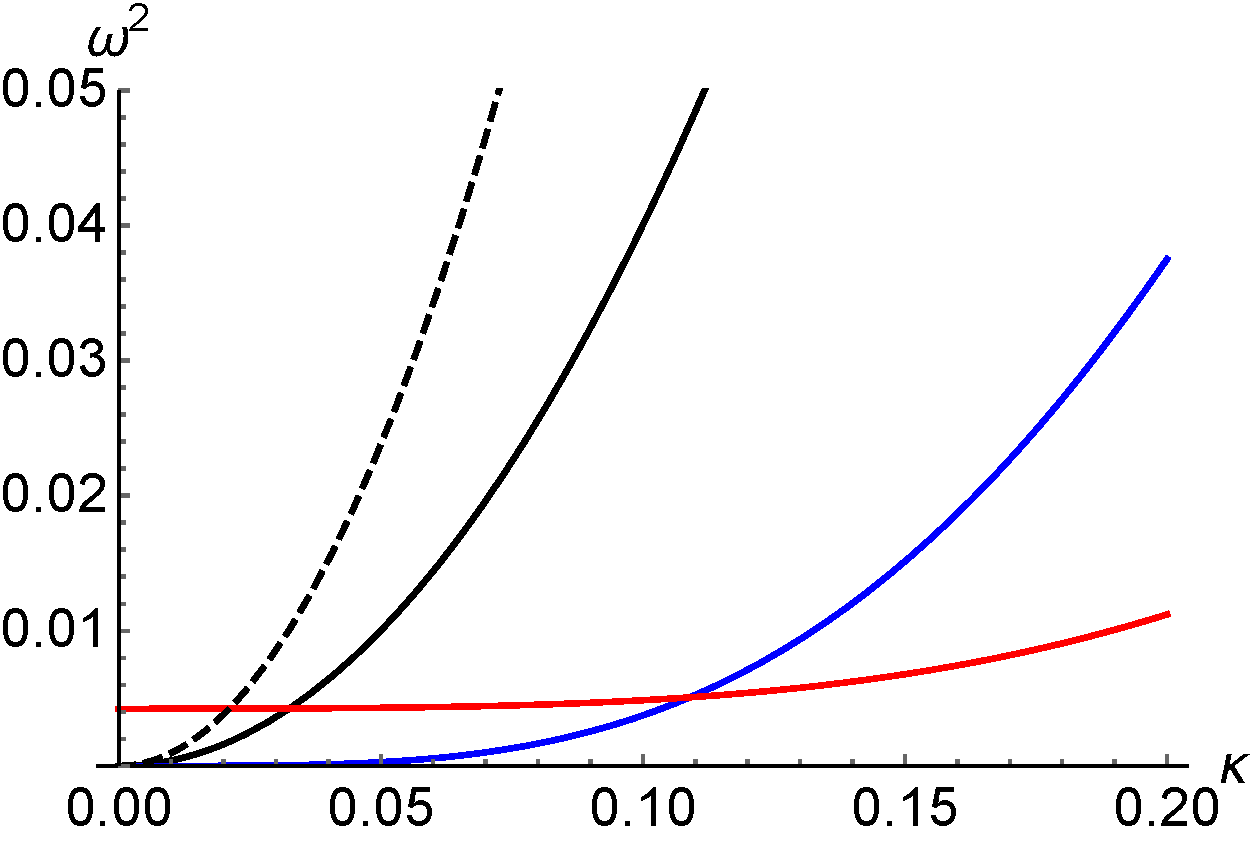}
\caption{(Color online) The low-frequency dispersion relations for CNT vibrations corresponding to various circumferential wave number $n$. Black, red and blue curves correspond to $n=0$, $n=1$ and $n=2$, respectively.}
}
\label{fig:spectr0}
\end{figure}

The problem of the carbon nanotubes (CNTs) oscillations is in the focus of various fields of  physics, nanoelectronics, chemistry, and biological sciences \cite{CLi03,Sazonova2004,Anantram06}.
These processes were investigated by different methods -- from \textit{ab initio} quantum studies up to the methods of thin elastic shells theory \cite{Mahan02,CYWang04,Silvestre11}.
The latter is especially interesting from the viewpoint of the occurrence of the nonlinear dynamical effects, which may be responsible for such phenomena as the anomalous thermoconductivity of the CNTs \cite{Berber00,Mingo05,Wang06} as well as the energy exchange and localization \cite{Smirnov2014,Smirnov2016PhysD}.
In our previous studies we demonstrated the possibility of the energy localization, which results from the resonant interaction of the nonlinear normal modes (NNMs) corresponding to the circumferential flexure branch \cite{Smirnov2014,Smirnov2016PhysD}.
The latter is the most low-frequency optical-type vibration branch of the spectrum with the gap frequency $\simeq 20 cm^{-1}$.
The specific frequency crowding near the long wavelength edge of the spectrum leads to the efficient interaction of the NNMs and, as a consequence, to appearance of the energy localization.
However, the resonant interaction is possible not only for the NNMs, which belong to the same oscillation branch.
In  figure \ref{fig:spectr0} the low frequency part of the spectrum is shown for the CNT with relative length $L/R=30$ ($L$ and $R$ are the CNT's length and its radius, respectively).
One can see that the frequencies of the beam-like and circumferential oscillations are extremely close for the longitudinal wave number $\kappa=3 \pi /L$.

In the framework of the linear theory these NNMs are independent and therefore, no interactions between them can exist.
In order to reveal the mode interaction and the effects, which can arise as its results, we need in the transition to the nonlinear vibration theory.
In this work we consider the CNT oscillations in the framework of the nonlinear Sanders-Koiter theory \cite{Amabili08}.
We demonstrate that the effective reduction of the equations of motion in the combination with the asymptotic analysis allows to study the nonlinear mode coupling and to reveal new stationary oscillations, which are absent in the framework of the linear approach, as well as to describe the non-stationary dynamics under condition of the 1:1 resonance.

\section{The model}\label{Model}

The dimensionless energy of the elastic CNT deformation can be written as follows:

\begin{eqnarray}\label{eq:Elastic_en}
E_{el} =\frac{1}{2} \int\limits_{0}^{1} \int\limits_{0}^{2 \pi}\left(\varepsilon_{\xi}^{2} + \varepsilon_{\varphi}^{2} + 2 \nu \varepsilon_{\xi} \varepsilon_{\varphi} + \frac{1 - \nu}{2} \varepsilon_{\xi \varphi}^{2}\right)d\xi d\varphi +\\
+\frac{\beta^{2}}{24} \int\limits_{0}^{1} \int\limits_{0}^{2 \pi}\left(\kappa_{\xi}^{2} + \kappa_{\varphi}^{2} + 2 \nu \kappa_{\xi} \kappa_{\varphi} + \frac{1 - \nu}{2} \kappa_{\xi \varphi}^{2}\right)d\xi d\varphi , \nonumber
\end{eqnarray}
where $\xi$ and $\varphi$ are the dimensionless coordinate along the CNT's axes ($\xi=x/L$) and azimuthal angle; $\varepsilon_{\xi}$, $\varepsilon_{\varphi}$ and $\varepsilon_{\xi\varphi}$ are the longitudinal, tangential and shear deformations, respectively. The variables $\kappa_{\xi}$, $\kappa_{\varphi}$ and $\kappa_{\xi\varphi}$ are the longitudinal, circumferential curvatures  and the torsion. The energy $E_{el}$ and the time $t$ are measured in the units $E_{0}=YRLh/(1-\nu^{2})$ and $t_{0}=1/\sqrt{Y/\rho R^{2}(1-\nu^{2})}$, respectively, with the Young modulus $Y$, the effective thickness $h$ and the Poisson ratio $\nu$. Two dimensionless parameters $\alpha=R/L$ and $\beta=h/R$ are essential for the further consideration.
 
In the framework of the Sanders-Koiter theory the middle surface deformations and the curvatures are written in the next form:

\begin{eqnarray}\label{eq:deformation}
\varepsilon_{\xi} = \alpha \frac{\partial u}{\partial \xi} + \frac{\alpha^{2}}{2}( \frac{\partial w}{\partial \xi})^{2} +\frac{1}{8}(\alpha \frac{\partial v}{\partial \xi}-\frac{\partial u}{\partial \varphi})^{2}, \nonumber \\ 
\varepsilon_{\varphi} = \frac{\partial v}{\partial \varphi} + w + \frac{1}{2} (\frac{\partial w}{\partial \varphi} -v)^{2}+\frac{1}{8}(\frac{\partial u}{\partial \varphi}-\alpha \frac{\partial v}{\partial \xi})^{2},   \\ 
 \varepsilon_{\xi \varphi} = \frac{\partial u}{\partial \varphi} + \alpha \frac{\partial v}{\partial \xi} +\alpha \frac{\partial w}{\partial \xi}(\frac{\partial w}{\partial \varphi}-v).  \nonumber 
\end{eqnarray}

\begin{eqnarray}\label{eq:curvation}
\kappa_{\xi} = - \alpha^{2} \beta \frac{\partial^{2} w}{\partial \xi^{2}}, \nonumber  \\ 
\kappa_{\varphi} = \beta \left(\frac{\partial v}{\partial \varphi} - \frac{\partial^{2} w}{\partial \varphi^{2}} \right), \\  
\kappa_{\xi \varphi} = \beta \left( - 2 \alpha \frac{\partial^{2} w}{\partial \xi \partial \varphi} + \frac{3 \alpha}{2} \frac{\partial v}{\partial \xi} - \frac{1}{2} \frac{\partial u}{\partial \varphi}\right), \nonumber
\end{eqnarray}
where $u$, $v$ and $w$ describe longitudinal, tangential and radial displacement fields, respectively.
Using expressions (\ref{eq:deformation} - \ref{eq:curvation}) one can variate elastic energy functional (\ref{eq:Elastic_en}) with respect to the displacements and obtain corresponding equations of motion.
When considering the linearized problem for simple-supported CNT one can represent the displacements as:

\begin{eqnarray}\label{eq:displacement}
u(\xi, \varphi, t)=u(\xi, t) \cos{n \varphi},  \nonumber \\
v(\xi, \varphi, t)=v(\xi, t) \sin{n \varphi}, \\
w(\xi, \varphi, t)=w(\xi, t) \cos{n \varphi}, \nonumber
\end{eqnarray} 
where $n$ is the azimuthal wave number, which can takes the integer values $n=0,1,2 \dots$.
Taking into account relations (\ref{eq:displacement}) one can estimate the dispersion relations for different values of $n$ using the linearised approximation of the equations of motion mentioned above.
The dispersion relations, which are showed in figure \ref{fig:spectr0} have been obtained for the values of azimuthal wave number $n=1$ (beam-like oscillations or BLO) and $n=2$ (circumferential flexure oscillations or CFO).

However, it is obvious that the full nonlinear set of the equations of motion leads, generally speaking, to unsolvable problem.
Therefore, in order to obtain any meaningful results, one should make the physically sensible hypotheses concerning to the shell displacement fields.
The specific features of the considered oscillations - BLO as well as CFO, are the smallness of the circumferential  ($\varepsilon_{\varphi}$) and the shear  ($\varepsilon_{\xi \varphi}$) deformations, in spite of that the displacements, which are included in them may be not small.
These assumptions can be written as follows:

\begin{eqnarray}\label{eq:hypotheses}
\varepsilon_{\xi \varphi} =0; \quad \varepsilon_{\varphi}=0.
\end{eqnarray}

In order to consider the interaction of the oscillations with the different azimuthal wave number, one should rewrite displacements (\ref{eq:displacement}) as a combination of the partial components:

\begin{eqnarray}\label{eq:displacement2}
u(\xi,\varphi,t)=   U_{0}(\xi,t)+U_{1}(\xi,t)\cos{n_{1}\varphi}+U_{2}(\xi,t) \cos{n_{2} \varphi}, \nonumber \\
v(\xi,\varphi,t)=   V_{1}(\xi,t)\sin{n_{1}\varphi}+V_{2}(\xi,t) \sin{n_{2} \varphi}, \\
u(\xi,\varphi,t)=   W_{0}(\xi,t)+W_{1}(\xi,t)\cos{n_{1}\varphi}+W_{2}(\xi,t) \cos{n_{2} \varphi}, \nonumber
\end{eqnarray}
where the functions ($U_{1}(\xi, t), V_{1}(\xi, t), W_{1}(\xi, t)$) and ($U_{2}(\xi, t), V_{2}(\xi, t), W_{2}(\xi, t)$) describe the BLO and CFO, respectively.
In spite of that we do not consider the axisymmetrical oscillations with $n=0$, the respective adding arises in the nonlinear elastic problem.

Taking into account hypotheses (\ref{eq:hypotheses}) and expressions (\ref{eq:deformation}) we have a possibility to define the relationships between longitudinal, tangential and radial components of the displacements:
\begin{eqnarray}\label{eq:uvexclusion}
V_{i}(\xi ,t) = -\frac{W_{i}(\xi ,t)}{n_{i}}, \quad 
U_{i}(\xi ,t) =  -\frac{\alpha}{n_{i}^2} \frac{\partial W_{i}(\xi ,t)}{\partial \xi}, \quad  i=1,2  \nonumber  \\ 
W_{0}(\xi ,t) = -\frac{1}{4} \sum_{i=1,2} \frac{1}{n_{i}^{2}} \left( \left(n_{i}^{2}-1 \right)^{2} W_{i}^{2}(\xi,t)  + \alpha^{2} \left( \frac{\partial W_{i}(\xi,t)}{\partial \xi} \right)^{2} \right) \nonumber  \\
\frac{\partial U_{0}(\xi,t)}{\partial \xi}= - \frac{\alpha}{4} \Bigl( \frac{n_{1}^{2}
+1 }{n_{1}^{2}} \left( \frac{\partial W_{1}(\xi,t)}{\partial \xi} \right)^{2} + \frac{n_{2}^{2}
+1 }{n_{2}^{2}} \left( \frac{\partial W_{2}(\xi,t)}{\partial \xi} \right)^{2} + \\
\frac{1}{2} \left( \frac{ (n_{1}^{2}-1)^{2}}{n_{1}^{4}} W_{1}(\xi, t)  \frac{\partial W_{1}(\xi,t)}{\partial \xi} + \frac{(n_{2}^{2}-1)^{2}}{n_{2}^{4}} W_{2}(\xi,t) \frac{\partial W_{2}(\xi,t)}{\partial \xi}  \right)^{2} \Bigr), \nonumber
\end{eqnarray}


Putting  expressions (\ref{eq:displacement2}) with accounting (\ref{eq:uvexclusion}) into (\ref{eq:Elastic_en}), one can get the equations of motion in the next form:

\begin{eqnarray}\label{eq:eqW1}
\frac{\partial^{2} W_{1}}{\partial t^{2}}   -\frac{\alpha^2}{2} \frac{\partial^{4} W_{1}}{\partial t^{2} \partial \xi^{2}}+\alpha^{4} \frac{12+\beta^{2}}{24} \frac{\partial^{4} W_{1}}{\partial \xi^{4}}- \nonumber  \\\frac{9 \alpha^2}{16} \frac{\partial }{\partial \xi} \left( \frac{\partial W_{1}}{\partial \xi} \frac{\partial }{\partial t} \left( W_{2} \frac{\partial W_{2}}{\partial t} \right) \right) =0 \nonumber  \\
\frac{\partial^{2} W_{2}}{\partial t^{2}}   +\frac{3 \beta^{2}}{5} W_{2} - \frac{\alpha^2 \beta^{2} (3+\nu)}{10} \frac{\partial^2 W_{2}}{\partial \xi^2}-\frac{\alpha^4}{20} \frac{\partial ^{4} W_{2}}{\partial \xi^{2} \partial t^2}+  \nonumber  \\  \frac{\alpha^{4} (3+4\beta^2)}{60}\frac{\partial ^{4} W_{2}}{\partial \xi^{4}} + 
\frac{81}{40}W_{2}    \frac{\partial }{\partial t} \left( W_{2} \frac{\partial W_{2}}{\partial t} \right)+   \nonumber  \\  \frac{9 \alpha^{2}}{40} \Bigl[ W_{2} \left( \left(\frac{\partial ^2 W_{2}}{\partial \xi \partial t} \right) ^{2} +2 \frac{\partial^2}{\partial t^2} \left( \frac{\partial W_{1}}{\partial t} \right)^2   \right)  -   \nonumber  \\   
\frac{\partial^2 W_{2}}{\partial t^2}   \frac{\partial}{\partial \xi} \left( W_2 \frac{\partial W_{2}}{\partial \xi} \right) -\frac{\partial }{\partial \xi} \left( \frac{\partial W_{2}}{\partial \xi} \left(\frac{\partial W_{2}}{\partial t} \right)^2 \right)   \Bigr] =0
\end{eqnarray}

/The azimuthal wave numbers ($n_1=1, n_2 =2$) have been taken in order to simplify equations (\ref{eq:eqW1})./
It is easy to see that the linearization of equations (\ref{eq:eqW1}) leads to the uncoupled equations of the BLOs and CFOs.
In such a case we can estimate the dispersion relations for boundary conditions, corresponding to the simple-supported edges:

\begin{eqnarray}\label{eq:dsipersion1}
\omega_{1}^{2}   =\frac{\left(12+\beta ^2\right) }{12 \left(2+\alpha^{2} \kappa ^2 \right)} \alpha^{4}\kappa ^4 \\
\omega _2^2   =\frac{1}{20+\alpha^2 \kappa ^2}\left( 12 \beta ^2+2 \beta ^2 \left( 3+ \nu \right) \alpha^2 \kappa^2 +  \frac{\left(3+4 \beta ^2 \right)}{3} \alpha^4 \kappa^4 \right), \nonumber
\end{eqnarray}

where the longitudinal wave number $\kappa= \pi k$ with integer value $k$ specifies the number of half-wavelengths along the CNT axis ($k  = 0,1\dots$).

\begin{figure}
\centering{
\includegraphics[width=50mm]{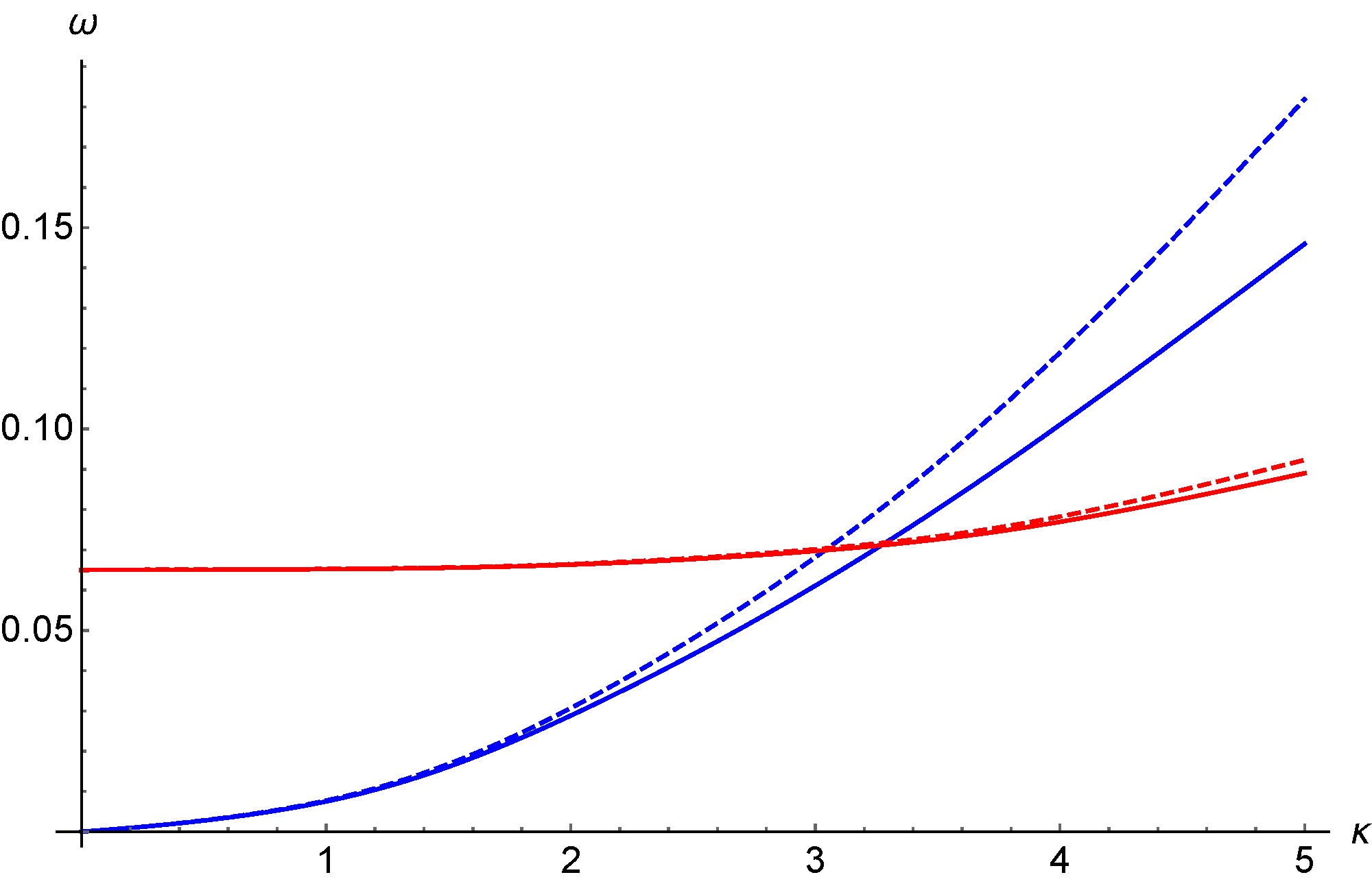}
\caption{(Color online) Dispersion relations for BLOs and CFOs. Blue and red curves corresponds to the azimuthal wave numbers $n_1=1$ and $n_2=2$, respectively. Solid and dashed curves show the exact values and the values estimated by equations (\ref{eq:dsipersion1}).}
}
\label{fig:DScompare}
\end{figure}

Figure \ref{fig:DScompare} shows the dispersion curves (\ref{eq:dsipersion1}) in comparison with the exact ones, which were estimated by the solution of the full linearized system.

One can see that the correspondence is good enough in the resonant frequency range.

The goal of our study is revealing the effects of the nonlinear coupling between resonantly interacting BLOs and CFOs.
It is obvious that equations (\ref{eq:eqW1}) can not be analyzed directly.
The semi-inverse method, which have been developed by us \cite{Kharkov2016,Smirnov2017}, will be discussed in next section.
 
\section{The semi-inverse method of analysis}\label{Method}
\subsection{Stationary solutions}\label{Stat}
The semi-inverse method for the analysis of the complex nonlinear systems was used for the investigation of the dynamics of the discrete nonlinear lattices \cite{Kharkov2016,CAP2016,Smirnov2017}, the forced oscillations of pendulum \cite{CAP2016_2,Kharkov2016} and, in slightly simplified reduction, to the study of the CNT oscillations  \cite{Smirnov2014,Smirnov2016PhysD,IJNM2017}.
The basis of this method consists in the presentation of the problem in terms of the complex variables with further analysis by the multiscale expansion method.
The feature of some dynamical problems is that the small parameter, which is required for the separation of the time scales, does not present in the initial formulation of the equations.
In such a case, this parameter is defined in the processes of the solution.
The stationary problem in the framework of the semi-inverse method is somewhat similar to the harmonic balance method \cite{Mickens2010}, but due to the presentation in terms of the complex variables, it turns out more simple and clear.
There is an additional advantage of the developed procedure.
The complex variables are the classical analogues of the creation and annihilation operators in the formalism of the secondary quantization.
In several cases, it may be useful for the comparison of the  classical and quantum mechanical problems.

Let us define new variables for the description of the CNT dynamics as follows:

\begin{eqnarray}\label{eq:w2psi}
\Psi_{j}(\xi, t)=\frac{1}{\sqrt{2}} \left( \frac{1}{\sqrt{\omega}} \frac{\partial W_{j}(\xi, t)}{\partial t}+ i \sqrt{\omega} \,W_{j}(\xi, t) \right), \quad j=1,2
\end{eqnarray}
The inverse transformation to the initial variables $W$ is written as:

\begin{eqnarray}\label{eq:psi2w}
W_{j}(\xi, t)   =\frac{-i}{\sqrt{2 \omega}} \left( \Psi_{j}(\xi, t)-\Psi_{j}^{*}(\xi, t) \right),  \nonumber  \\
\frac{\partial W_{j}(\xi, t)}{\partial t}   =\sqrt{\frac{\omega}{2}} \left( \Psi_{j}(\xi, t)+\Psi_{j}^{*}(\xi, t) \right), \quad j=1,\,2
\end{eqnarray}
where $\omega$ is an (yet) undefined frequency and the asterisk means the complex conjugation.
First of all we try to find the stationary solutions for equations (\ref{eq:eqW1}), which  correspond to the nonlinear normal modes (NNMs) for the coupled BLOs and CFOs.
Substituting relations (\ref{eq:w2psi}) into equations (\ref{eq:eqW1}) one can find the stationary one-frequency solution in next form:

\begin{eqnarray}\label{eq:statsol}
\Psi_{j}(\xi,t)= \psi_{j}(\xi) e^{i \omega t}, \quad j=1,\,2
\end{eqnarray}
where functions $\psi_{j}$ do not depend on the time.

As a result, we obtain the equations which have to determine the profiles of the NNMs:

\begin{eqnarray}\label{eq:eqstat}
-   \frac{\omega}{2} \psi_{1}+\alpha^2\frac{\omega}{4} \frac{\partial^2 \psi_{1}}{\partial \xi^2} + \alpha^{4} \frac{12+\beta^2}{48 \omega} \frac{\partial^4 \psi_{1}}{\partial \xi^4}+ \frac{9 \alpha^2}{32} \frac{\partial }{\partial \xi} \left( \psi_{2}^{2} \frac{\partial \psi_{1}^{*}}{\partial \xi} \right) =0,   \nonumber   \\
-   \frac{5 \omega^2-3 \beta^2}{10 \omega} \psi_{2}+ \alpha^2 \frac{\omega^2-2\beta^2 (3+\nu)}{40 \omega} \frac{\partial^2 \psi_{2}}{\partial \xi^2}+  \\  
\alpha^4 \frac{3+4\beta^2}{120 \omega} \frac{\partial^4 \psi_{2}}{\partial \xi^4} 
-   \frac{81}{80}|\psi_{2}|^2 \psi_{2}+  \nonumber  \\
\frac{9 \alpha^2}{80}  \left( \frac{\partial }{\partial \xi} \left( \psi_{2}^2 \frac{\partial \psi_{2}^{*}}{\partial \xi} \right) - \psi_{2}^{*} \left( \left(\frac{\partial \psi_{2}}{\partial \xi} \right)^2 - 4 \left( \frac{\partial \psi_{1}}{\partial \xi} \right)^2 \right) \right) =0,  \nonumber
\end{eqnarray}

It is easy to see that functions

\begin{eqnarray}\label{eq:psistat}
\psi_{j}=\sqrt{X_{j}} e^{i \kappa \xi}
\end{eqnarray}
are the solutions of equations (\ref{eq:eqstat}), if the relations

\begin{eqnarray}\label{eq:Astat}
\left(- \frac{\omega}{2}-\alpha^2 \frac{\omega}{4} \kappa^2 +\alpha^4 \frac{12+\beta^2}{48 \omega} \kappa^4 \right) \sqrt{X_{1}} +\frac{9 \alpha^2}{32} \kappa^2 \sqrt{X_{1}} X_{2} =0   \nonumber  \\
\frac{1}{120 \omega} \Bigl( 36 \beta^2 - 60 \omega^2 + 3 \alpha^2 \left( 2 \beta^2 (3+\nu) - \omega^2 \right) \kappa^2 +   \\
\alpha^4 (3+4 \beta^2) \kappa^4 \Bigr) \sqrt{X_{2}}-  \frac{9}{80}  \left( 9- 2 \alpha^2 \kappa^2 \right) X_{2}^{3/2}+ \frac{9 \alpha^2}{20} \kappa^2 X_{1} \sqrt{X_{2}} = 0 \nonumber
\end{eqnarray}
are satisfied.

The relations between the frequency and amplitudes of the BLOs and CFOs can be obtained due to the simple coupling between the modules of functions $|\Psi_{j}|^2=X_{j}$ and partial amplitudes $A_{j}$:

\begin{eqnarray}\label{eq:modulus}
X_{j}=\frac{\omega}{2} A_{j}^{2}
\end{eqnarray}
which can be seen from the definition (\ref{eq:psi2w}) of the functions $\Psi_{j}$.

Using relations (\ref{eq:modulus}), the following expressions for the frequencies of the stationary nonlinear coupling oscillations can be obtained:

\begin{eqnarray}\label{eq:nonlinfrequency}
\omega_{1}^2   = \frac{4 \alpha ^4 \left(12+\beta ^2\right) \kappa ^4}{96+48 \alpha ^2 \kappa ^2-27 \alpha ^2 \kappa ^2 A_2^2 } \nonumber  \\
\omega_{2}^2   = \frac{4 \left(36 \beta ^2+6 \alpha ^2 \beta ^2  (\nu +3) \kappa ^2+\alpha ^4 \left(4 \beta ^2+3\right) \kappa ^4 \right)}{3 \left( 4 (20+ \alpha ^2 \kappa^2) -36 \alpha ^2  \kappa ^2 A_1^2+9  \left(9-2 \alpha ^2 \kappa ^2\right) A_2^2 \right)}.
\end{eqnarray}

\begin{figure}
\centering{
\includegraphics[width=50mm]{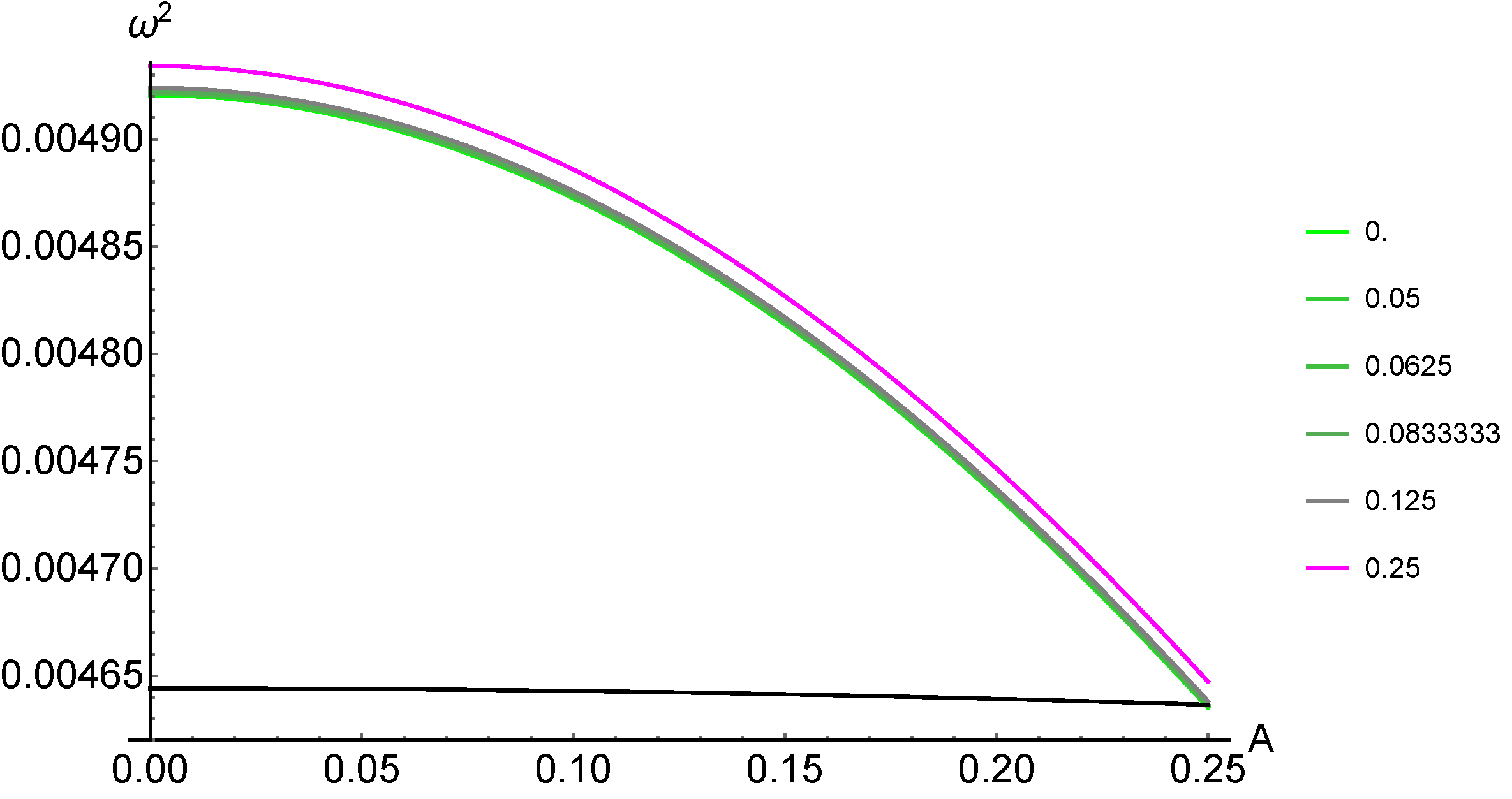}
\caption{(Color online) Dependence of the frequencies $\omega_{1}$ (black) and $\omega_{2}$ (color curves) accordingly (\ref{eq:nonlinfrequency}) on the amplitude of CFOs  at different BLOs' amplitudes for the CNT with $\alpha=1/30$. The values of BLOs amplitudes are shown on the right of figure. The wave number $\kappa=3\pi$.}
}
\label{fig:resonant1}
\end{figure}

One can see, that equations (\ref{eq:nonlinfrequency}) coincide with dispersion relations (\ref{eq:dsipersion1}) in the small-amplitudes limit $A_{1} \to 0$ and $A_{2} \to 0$.
One should notice that equations (\ref{eq:nonlinfrequency}) are valid only for resonant interaction of the modes, i.e. at $\omega_{1} \simeq \omega_{2}$, due to that we find the solution for the single-frequency motion (\ref{eq:statsol}).
Figure \ref{fig:resonant1} shows the dependences of the frequencies $\omega_{1}$, $\omega_{2}$ on the CFOs amplitude with the various BLOs amplitude at the "resonant" wave number $\kappa=3 \pi$.

%

\subsection{Multi-scale expansion}

The analysis of the non-stationary solutions of equations (\ref{eq:eqW1}) needs in the method, which allows to describe the time evolution of the BLOs and CFOs in the vicinity of the resonance.
This objective may be realized by the multi-scale expansion method, which consists in the separation of the time scales.
Such a procedure requires a small parameter, the value of which can be estimated with the parameters of the problem.
It is obvious that the time separation is better with a smaller value of this parameter.

We consider the "carrier" time $t$ in equation (\ref{eq:statsol}) as the basic "fast" time.
Let us define the value "$\varepsilon$" as a small parameter.
Its magnitude will be estimated further.
We introduce the time hierarchy as follows:

\begin{equation}
\tau_{0}=t, \, \tau_{1}=\varepsilon \tau_{0}, \, \tau_{2}=\varepsilon^2 \tau_{0}, \dots
\end{equation}
and
\begin{equation}
\frac{d}{d\,t}=\frac{\partial }{\partial \tau_{0}}+\varepsilon \frac{\partial }{\partial \tau_{1}}+\varepsilon^2 \frac{\partial }{\partial \tau_{2}} + \dots
\end{equation}

Keeping in mind that the oscillations of the shell should be small enough, we represent the functions $\Psi_{j}$ as:

\begin{eqnarray}\label{eq:series1}
\Psi_{j}(\xi, t)=\varepsilon \bigl( \psi_{j,0}(\xi, \tau_{1},\tau_{2})+\varepsilon \psi_{j,1}(\xi, \tau_{1},\tau_{2})+  \nonumber  \\
\varepsilon^2 \psi_{j,2}(\xi, \tau_{1},\tau_{2})+ \dots \bigr) e^{i \omega \tau_{0}}
\end{eqnarray}

Substituting expansions (\ref{eq:series1}) into equations (\ref{eq:eqW1}) and performing the averaging with respect to the "fast" time $\tau_{0}$, we obtain the equations for the functions $\psi_{j,n}$ ($j=1,2$, $n=0,1,\dots$) in the different orders of small parameter.
Because the respective procedure is  well known and has been described many times  (see, in particular, \cite{Smirnov2016PhysD,VVS2010}), we omit the details. 

$\varepsilon^1$
\begin{eqnarray}\label{eq:E1}
-  \frac{\omega}{2}   \psi_{1,0}+\frac{\alpha^2 \omega }{4} \frac{\partial^2 \psi_{1,0}}{\partial \xi^2}+\frac{\alpha^4 (12+\beta^2)}{48\omega} \frac{\partial^4 \psi_{1,0}}{\partial \xi^4 }=0  \nonumber  \\
\left( -  \frac{\omega}{2}+\frac{3 \beta^2}{10 \omega} \right)   \psi_{2,0}+\frac{\alpha^2}{40} \left( \omega-\frac{2 \beta^2 (3+\nu)}{\omega} \right) \frac{\partial^2 \psi_{2,0}}{\partial \xi^2} +   \\
\frac{\alpha^4 (3+4 \beta^2)}{120 \omega} \frac{\partial^4 \psi_{2,0}}{\partial \xi^4} = 0 \nonumber 
\end{eqnarray}

One can  verify easily that the first (second) of  equations (\ref{eq:E1}) is satisfied exactly for the functions $\psi_{j,0} \sim \exp(i \kappa \xi)$, if  the frequency $\omega$ corresponds to  one of  dispersion relations (\ref{eq:dsipersion1}).
However, $\omega_{1}$ is equal to $\omega_{2}$ only approximately (see Fig. \ref{fig:resonant1}).
Therefore, when the values of $\omega_{1}$ and $\omega_{2}$ are close, equations (\ref{eq:E1}) are satisfied with some accuracy, which is determined by the frequencies detuning.
This detuning is the required small parameter, which is needed for the separation of the time scales.
In such a case, the expressions in equations (\ref{eq:E1}) have to be moved into equations of another order of $\varepsilon$.
As it will be shown further, the frequency detuning turns out to be of the second order with respect to the parameter $\varepsilon$.

$\varepsilon^2$
\begin{eqnarray}\label{eq:E2}
i \frac{\partial \psi_{1,0}}{\partial \tau_{1}}   - i\frac{\alpha^2}{2} \frac{\partial }{\partial \tau_{1}} \frac{\partial^2 \psi_{1,0}}{\partial \xi^2}-  \frac{\omega}{2}  \psi_{1,1}+\frac{\alpha^2 \omega }{4} \frac{\partial^2 \psi_{1,1}}{\partial \xi^2}+   \nonumber  \\\frac{\alpha^4 (12+\beta^2)}{48\omega} \frac{\partial^4 \psi_{1,1}}{\partial \xi^4 }=0  \\
i \frac{\partial \psi_{2,0}}{\partial \tau_{1}}   - i\frac{\alpha^2}{20} \frac{\partial }{\partial \tau_{1}} \frac{\partial^2 \psi_{2,0}}{\partial \xi^2} + \left( -  \frac{\omega}{2}+\frac{3 \beta^2}{10 \omega} \right)  \psi_{2,1}+  \nonumber   \\  \frac{\alpha^2}{40}   \left( \omega-\frac{2 \beta^2 (3+\nu)}{\omega} \right) \frac{\partial^2 \psi_{2,1}}{\partial \xi^2} +  \frac{\alpha^4 (3+4 \beta^2}{120 \omega} \frac{\partial^4 \psi_{2,1}}{\partial \xi^4} = 0   \nonumber
\end{eqnarray}

Spoken above about the oscillations frequencies and their detuning in equations (\ref{eq:E1}) is absolutely correct with respect to the expressions in equations (\ref{eq:E2}), which contain the functions of the first approximation $\psi_{j,1}$. 
Therefore, one should consider that these terms have to be moved from equations of this order of the small parameter.
So, equations (\ref{eq:E2}) can be written as follows:

\begin{eqnarray}\label{eq;E2x}
i \frac{\partial \psi_{1,0}}{\partial \tau_{1}}   - i\frac{\alpha^2}{2} \frac{\partial }{\partial \tau_{1}} \frac{\partial^2 \psi_{1,0}}{\partial \xi^2} =0  \nonumber   \\
i \frac{\partial \psi_{2,0}}{\partial \tau_{1}}   - i\frac{\alpha^2}{20} \frac{\partial }{\partial \tau_{1}} \frac{\partial^2 \psi_{2,0}}{\partial \xi^2}  = 0
\end{eqnarray}

These equations show that the functions $\psi_{1,0}$, $\psi_{2,0}$ do not depend on the time $\tau_{1}$.

The arguments which are similar to mentioned above should be applied to the equations of the next order by the small parameter.
According to the series (\ref{eq:series1}) the nonlinear terms should be included into equations of the third order of the small parameter.
The linear terms from equations (\ref{eq:E1}) have the same order due to our assumptions about detuning.
Omitting some tedious calculations, one can write the equations of the third order  as follows:

$\varepsilon^3$
\begin{eqnarray}\label{eq:E3}
i \frac{\partial \psi_{1,0}}{\partial \tau_{2}}  - i\frac{\alpha^2}{2} \frac{\partial }{\partial \tau_{2}} \frac{\partial^2 \psi_{1,0}}{\partial \xi^2}  -   \frac{\omega}{2}  \psi_{1,0}+\frac{\alpha^2 \omega }{4} \frac{\partial^2 \psi_{1,0}}{\partial \xi^2}+  \nonumber  \\
\frac{\alpha^4 (12+\beta^2)}{48\omega} \frac{\partial^4 \psi_{1,0}}{\partial \xi^4 } +\frac{9 \alpha^2}{32} \frac{\partial}{\partial \xi} \left( \psi_{2,0}^2 \frac{\partial \psi_{1,0}^{*}}{\partial \xi} \right)=0 \nonumber \\
i \frac{\partial \psi_{2,0}}{\partial \tau_{2}}  - i\frac{\alpha^2}{20} \frac{\partial }{\partial \tau_{2}} \frac{\partial^2 \psi_{2,0}}{\partial \xi^2} + \left( -  \frac{\omega}{2}+\frac{3 \beta^2}{10 \omega} \right)  \psi_{2,0}+  \quad \qquad  \\   \frac{\alpha^2}{40} \left( \omega-\frac{2 \beta^2 (3+\nu)}{\omega} \right) \frac{\partial^2 \psi_{2,0}}{\partial \xi^2} +  \frac{\alpha^4 (3+4 \beta^2}{120 \omega} \frac{\partial^4 \psi_{2,0}}{\partial \xi^4} - \frac{81}{80} |\psi_{2,0}|^2 \psi_{2,0}+ \nonumber  \\
\frac{9 \alpha^2}{80} \left( \frac{\partial}{\partial \xi} \left( \psi_{2,0}^2 \frac{\partial \psi_{2,0}^{*}}{\partial \xi} \right) - \left( \left( \frac{\partial \psi_{2,0}}{\partial \xi} \right)^2 +4 \left( \frac{\partial \psi_{1,0}}{\partial \xi} \right)^2 \right) \psi_{2,0}^{*} \right) = 0  \nonumber
\end{eqnarray}

So, we have obtained the evolution equations for the main approximation complex functions $\psi_{j,0}$ under conditions of their resonant interaction with frequencies detuning $\sim \varepsilon^2$.

One can notice that the second of equations (\ref{eq:E3}), which describes the evolution of circumferential flexure vibrations of the CNT slightly differs from the respective equation in \cite{Smirnov2016PhysD}.
The main reason is that the previous consideration was based on the another small parameter, which correlates with the gap between the modes' frequencies in the only branch.

One can easily see that  the solutions for equations (\ref{eq:E3}) are the plane-wave functions $exp( i \, \kappa \, \xi)$ with slowly variating amplitudes and we will consider their behaviour at various amplitudes in the next section.

\subsection{Analysis of the steady states solutions and non-stationary dynamics}

Let us introduce the new variables, which describe the evolution of the normal modes in the slow time $\tau_{2}$:

\begin{eqnarray}\label{eq:varphi}
\psi_{j,0}(\xi, \tau_{2})=\varphi_{j} (\tau_{2}) e^{i \,\kappa \, \xi}, \quad j=1,2
\end{eqnarray}

Substituting functions (\ref{eq:varphi}) into equations (\ref{eq:E3}), we obtain the set of ODEs for the functions $\varphi_{j}$:

\begin{eqnarray}\label{eq:eqphi1}
i   \left(1+\frac{\alpha^2 \kappa^2}{2} \right) \frac{\partial \varphi_{1}}{\partial \tau_2} - \left(-\frac{\omega}{2}+\alpha^2 \kappa^2 \omega - \alpha^4 \kappa^4 \frac{12+\beta^2}{48 \omega} \right) \varphi_{1} +  \nonumber  \\   \frac{9 \alpha^2 \kappa^2}{32} \varphi_{2}^2 \varphi_{1}^{*} =0  \nonumber  \\
i   \left(1+\frac{\alpha^2 \kappa^2}{20} \right) \frac{\partial \varphi_{2}}{\partial \tau_2} - \Bigl( \frac{5\omega^2-3 \beta^2}{10\omega}+\alpha^2 \kappa^2\frac{ (\omega^2 - 2\beta^2(3+\nu))}{40 \omega} \\
 -   \alpha^4 \kappa^4 \frac{3-4 \beta^2}{120 \omega} \Bigr) \varphi_{2} - \frac{9}{80} \left( 9-2 \alpha^2 \kappa^2 \right) | \varphi_{2} |^2 \varphi_{2} + \frac{9 \alpha^2 \kappa^2}{20} \varphi_{1}^2 \varphi_{2}^{*} = 0 \nonumber
\end{eqnarray}

In order to obtain the Hamilton system we need in the renormalization of the variables.
One can show that the functions

\begin{eqnarray}\label{eq:varchi}
\chi_{1}(\tau_2) = \varphi_{1}(\tau_2); \quad \chi_{2}(\tau_2)= \frac{4 \sqrt{2+\alpha^2 \kappa^2}}{\sqrt{20+\alpha^2 \kappa^2}} \varphi_{2}(\tau_2)
\end{eqnarray}
form the set of the canonical variables for the system with the Hamilton function

\begin{eqnarray}\label{eq:Hamiltonian1}
H = a_{1} |\chi_{1}|^2 + a_{2} |\chi_{2}|^2+b_{1} |\chi_{2}|^4 +b_{2} \left( \chi_{1}^2 \chi_{2}^{* \, 2}+\chi_{1}^{* \, 2} \chi_{2}^2 \right),
\end{eqnarray}
where
\begin{eqnarray}\label{eq:coeff1}
a_{1} =   \frac{-24 \omega ^2-12 \alpha ^2 \kappa ^2 \omega ^2+\alpha ^4  \kappa ^4 \left(\beta ^2+12\right)}{24 \omega  \left(2+\alpha ^2 \kappa ^2 \right)} \nonumber  \\
a_{2} =   \frac{36 \beta ^2-60 \omega ^2+3 \alpha ^2 \kappa ^2 \left(2 \beta ^2 (\nu +3)-\omega ^2\right)+\alpha ^4 \kappa ^4 \left(4 \beta ^2+3\right) }{6 \omega  \left(20+ \alpha ^2 \kappa ^2 \right)}  \\
b_{1} =    \frac{18 \left(-18-5 \alpha ^2 \kappa ^2 +2 \alpha ^4 \kappa ^4 \right)}{ \left(20+\alpha ^2 \kappa ^2 \right)^2} \nonumber \\
b_{2} =   \frac{9 \alpha ^2 \kappa ^2}{2 \left(20+\alpha ^2 \kappa ^2 \right)}  \nonumber
\end{eqnarray}

In such a case the equatons of motion can be written as:

\begin{equation}
i \frac{\partial \chi_{j}}{\partial \tau_{2}} = -\frac{\partial H}{\partial \chi_{j}^{*}}.
\end{equation}

One should notice that the respective equations of motion have an additional integral, besides the integral of the energy.
It is the integral, which is termed as the "occupation number" in the quantum mechanical problems.
In our case it is expressed as follows:

\begin{equation}\label{eq:occupation}
X = |\chi_{1}|^2+|\chi_{2}|^2.
\end{equation}

\begin{figure}
\centering{
(a)\includegraphics[width=40mm]{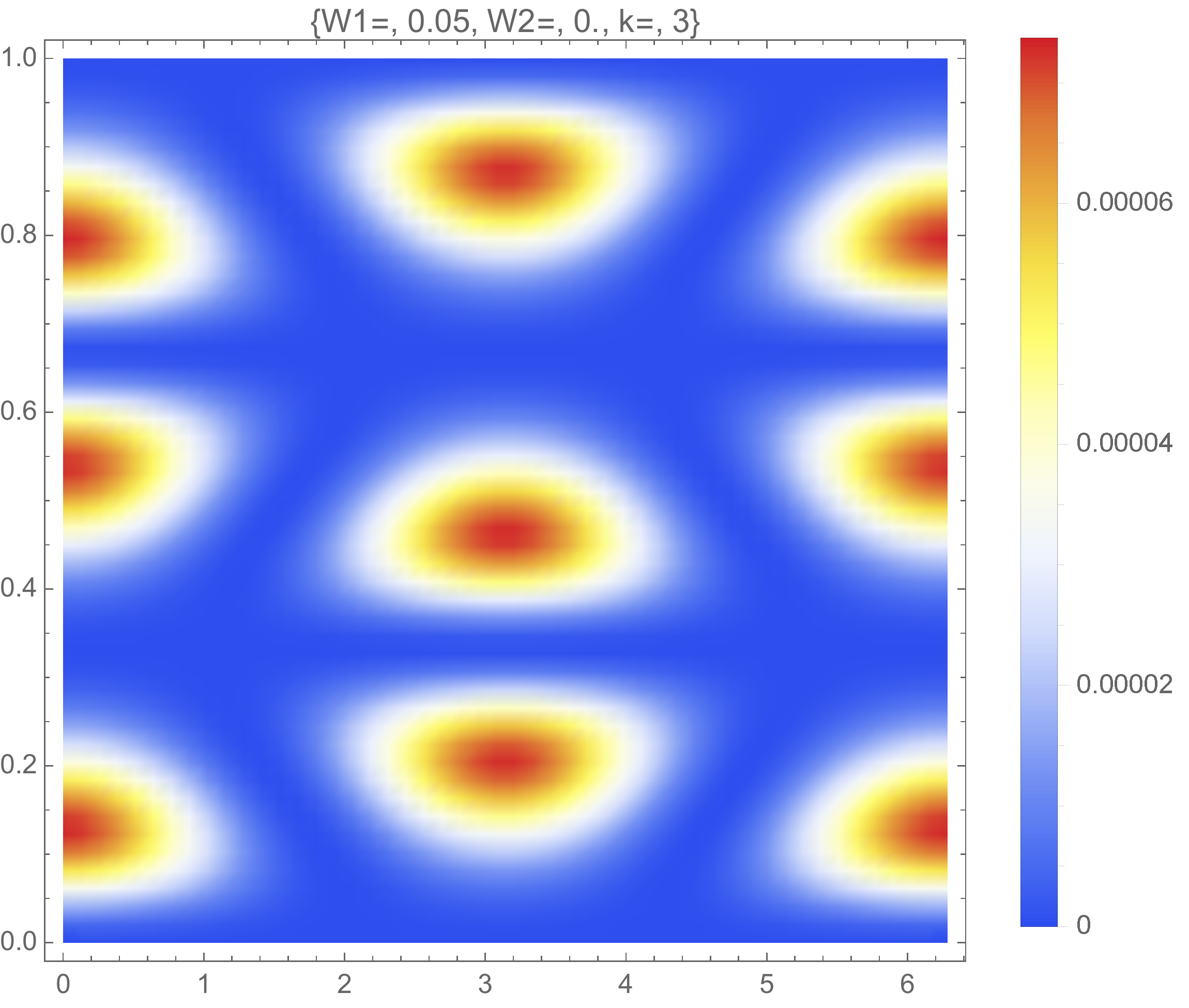} \quad (b) \includegraphics[width=40mm]{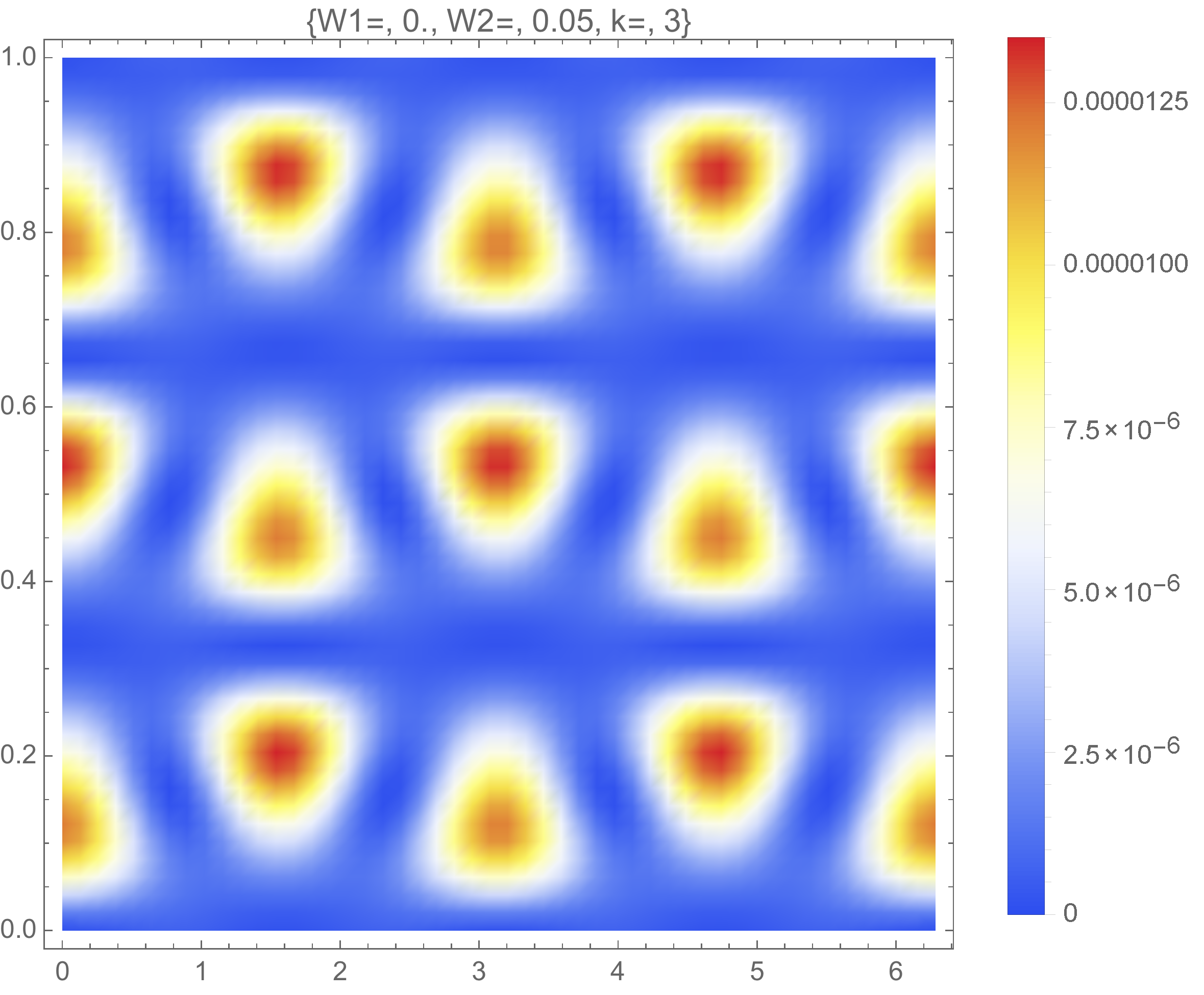} \\
(c)\includegraphics[width=40mm]{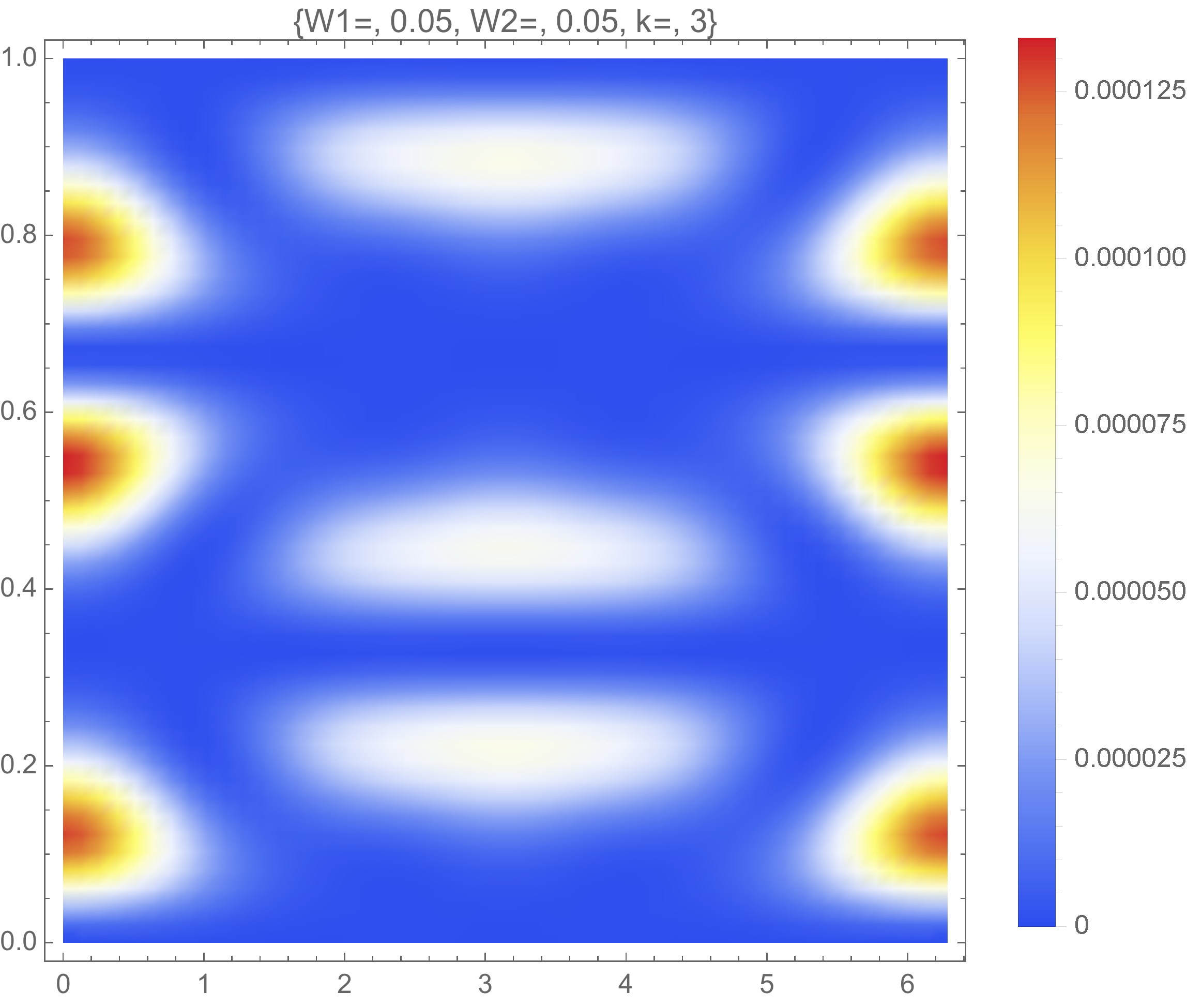} \quad (d) \includegraphics[width=40mm]{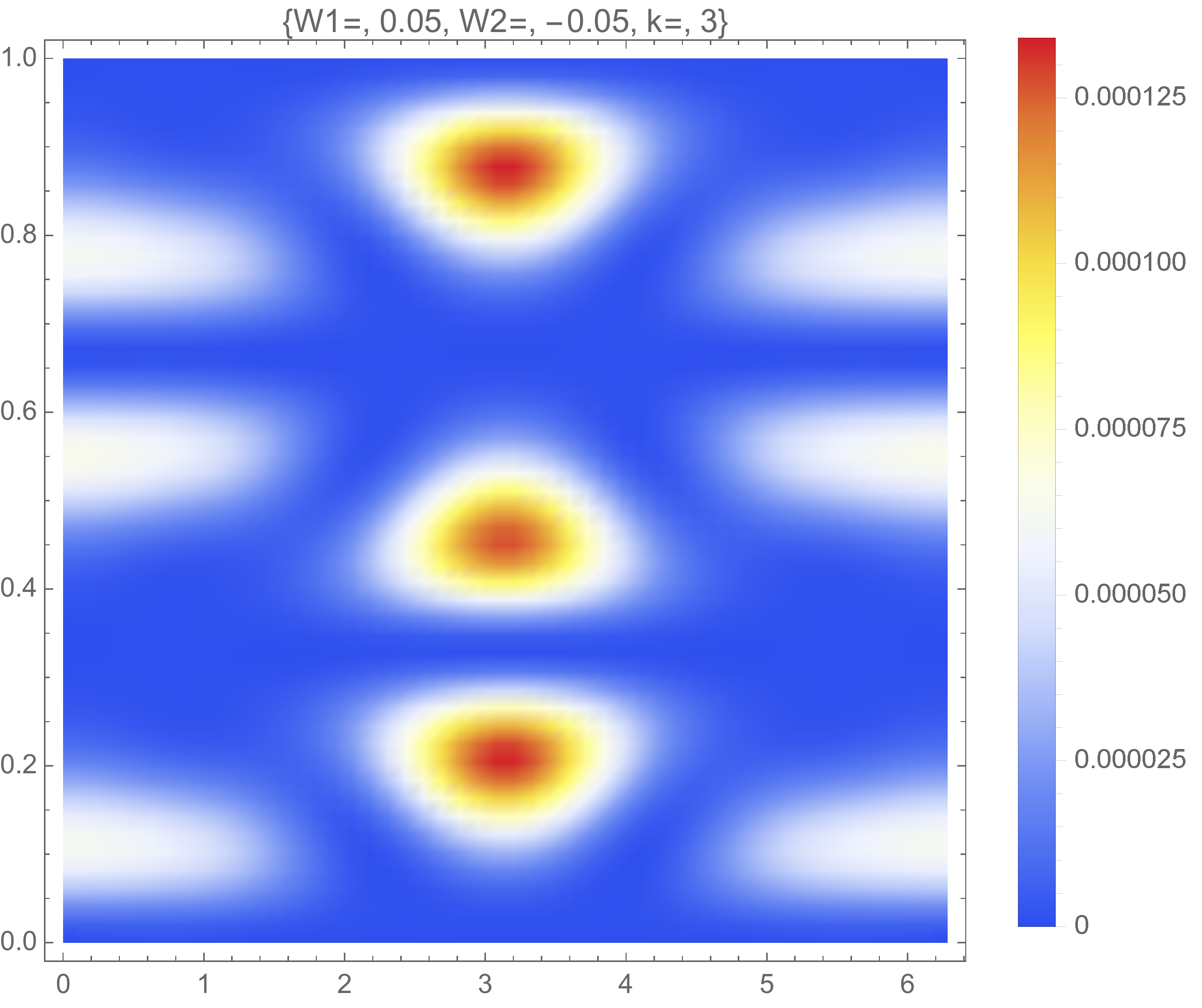}
\caption{(Color online) Elastic energy distribution on the surface of the CNT. (a) Beam-lke mode; (b) - circumferential flexure mode; (c) "sum" of BLO and CFO; (d) "difference" BLO and CFO. $\alpha =1/30, \kappa=3 \pi$; Amplitudes  $w=0.05$. }
}
\label{fig_energymap1}
\end{figure}

Before starting the analysis of the system with the Hamilton function (\ref{eq:Hamiltonian1}), one should discuss the question: what effects do result from the interaction of the BLOs and CFOs?
In order to answer this question let us consider the  elastic energy distribution corresponding to considered oscillations.
Figure~\ref{fig_energymap1} shows the "surface" energy distributions for the BLOs (\ref{fig_energymap1}a), the CFOs (\ref{fig_energymap1}b) and their combinations (\ref{fig_energymap1}c, d).

\begin{figure}
\centering{
(a)\includegraphics[width=40mm]{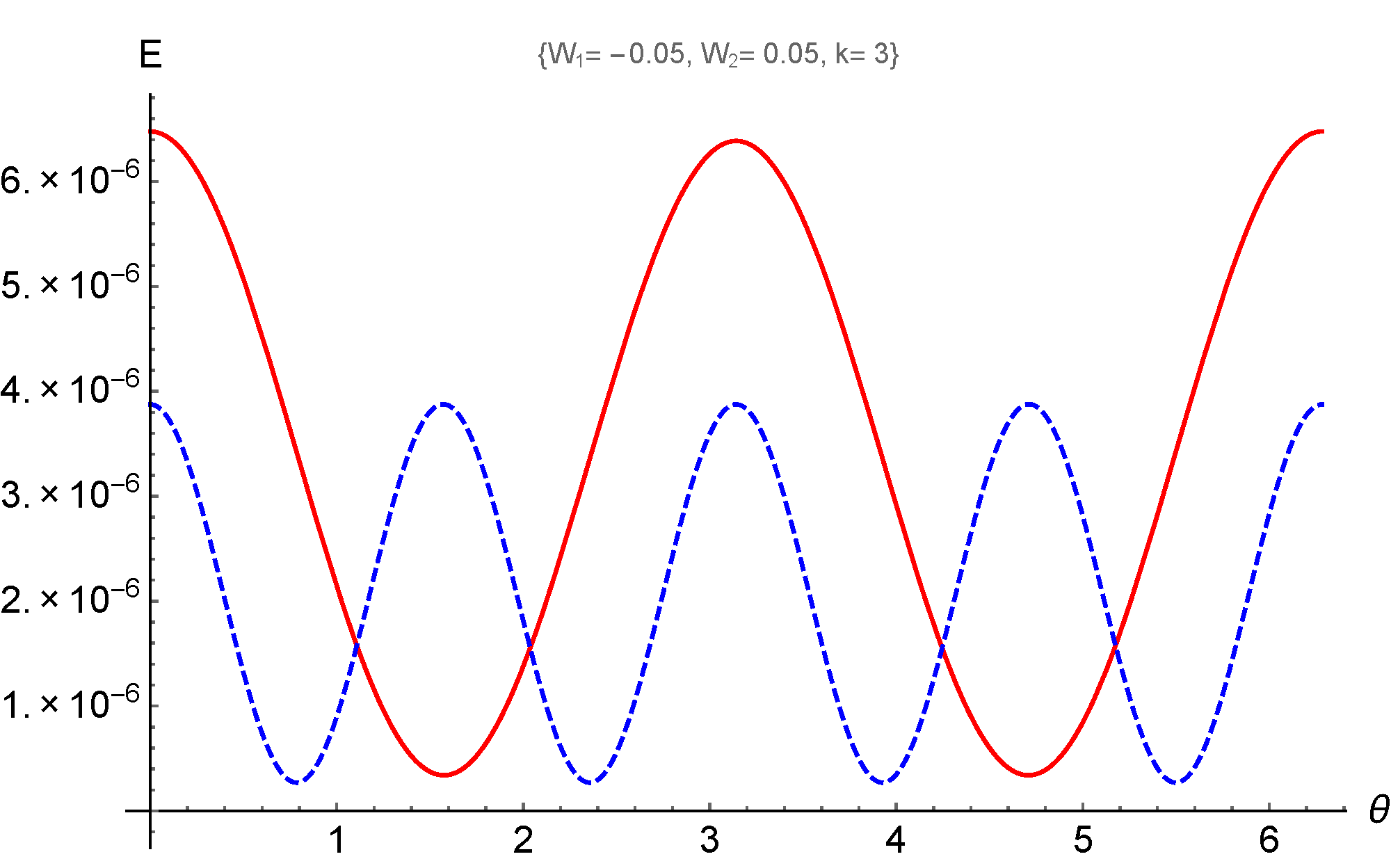} \quad (b) \includegraphics[width=40mm]{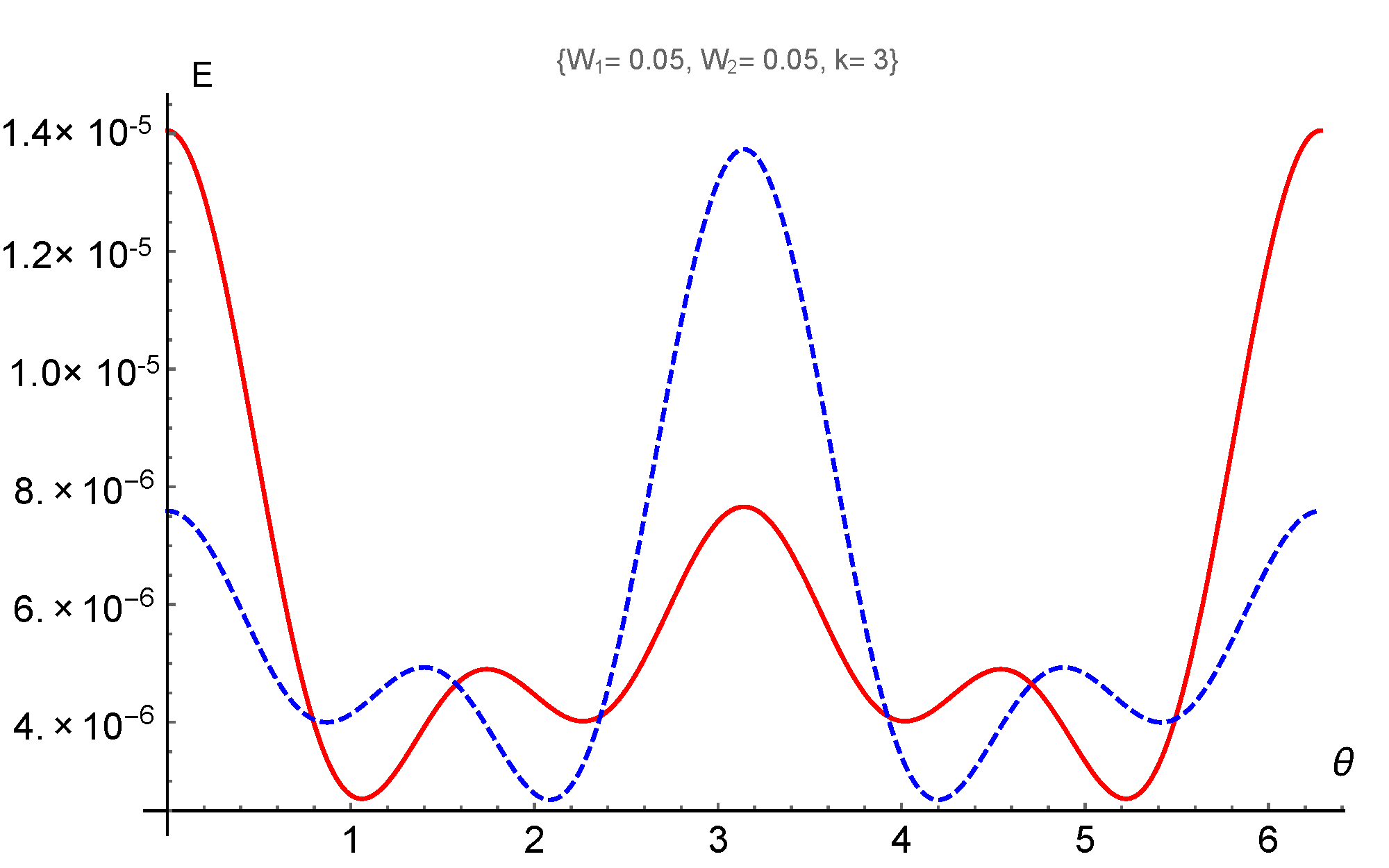}
\caption{(Color online) The energy distributions along the azimuthal coordinate. (a) Solid red and dashed blue curves correspond to BLO and CFO, respectively; (b) solid red and dashed blue curves correspond to "sum" and "difference" of BLO and CFO, respectively.}
}
\label{fig_energy1}
\end{figure}

The energy distributions along the azimuthal coordinate are shown in figure \ref{fig_energy1}. These curves have been obtained by integrating the distribution shown in Figs. \ref{fig_energymap1} along the longitudinal coordinate. 

In the case of non-interacting normal modes, due to a small difference between BLOs' and CFOs' frequencies, the transitions between the energy distributions depicted in Fig. \ref{fig_energymap1}(c) and  Fig. \ref{fig_energymap1}(d) are similar to the beating in the system of the  weakly-coupled linear oscillators.
However, the nonlinear coupling of the BLOs and CFOs may result in some other scenarios \cite{VVS2010,Smirnov2016PhysD}.
As it was shown \cite{VVS2010} the nonlinear normal modes do not represent the adequate notions under conditions of $1:1$ resonance.
It happens due to that the considered system (it may be a nonlinear lattice or a CNT) is separated into some domains with coordinate motion of its components, while the motion in the different domains differs essentially.
In such a case, the description of the system's dynamics in the terms of the domain coordinates is more appropriate \cite{CAP2016}.
For the system under consideration the domain coordinates correspond to the linear combination of the functions $\chi_{1}$ and $\chi_{2}$:

\begin{eqnarray}\label{eq:domains1}
\sigma_{1}=\frac{1}{\sqrt{2}} \left( \chi_{1}+ \chi_{2} \right); \quad \sigma_{2}=\frac{1}{\sqrt{2}} \left( \chi_{1} - \chi_{2} \right).
\end{eqnarray}

One can show that the relations $\sigma_{1} \gg\sigma_{2}$ and $\sigma_{2} \ll \sigma_{1}$ correspond to the energy distributions, which are depicted in Fig. \ref{fig_energymap1} (c) and (d), respectively.
Transformation (\ref{eq:domains1}) preserves integrals (\ref{eq:Hamiltonian1}, \ref{eq:occupation}).

It is convenient to introduce the polar representation of the variables $\sigma_{1}$ and $\sigma_{2}$.
Due to the presence of integral (\ref{eq:occupation}), one can reduce the phase space of the system for the fixed value of $X$:

\begin{eqnarray}\label{eq:polar}
\sigma_{1}=\sqrt{X} \cos{\theta} e^{i \delta_{1}}; \quad  \sigma_{2}=\sqrt{X} \sin{\theta} e^{i \delta_{2}}
\end{eqnarray}

It can be shown that the energy of the system turns out to be dependent on the difference of the phases $\Delta=\delta_{1}-\delta_{2}$ only.

\begin{eqnarray}\label{eq:hamiltonian2}
H   = \frac{1}{4} X \Bigl( 2 \left( a_{1}+a_{2} - \left(a_{1}-a_{2} \right)\sin{2 \theta} \cos{\Delta} \right) + \nonumber \\
  X   \left( b_{1} (1-\cos{\Delta} \sin{2 \theta} )^{2} + b_{2} \left( 4 \cos^{2}{\Delta} + \left( 2- \cos^{2}{\Delta} \right) \sin{4 \theta} \right) \right) \Bigr)
\end{eqnarray}

In such a case the phase space is two-dimensional one and its structure can be studied by the phase portrait method.

The equations of motion in the terms $\theta$ and $\Delta$ results from the relations:

\begin{eqnarray}\label{eq:eqangle}
\sin{2 \theta} \frac{\partial \theta}{\partial \tau_{2}} = - \frac{\partial H}{\partial \Delta}; \quad \sin{2 \theta} \frac{\partial \Delta}{\partial \tau_{2}} = \frac{\partial H}{\partial \theta}
\end{eqnarray}

\begin{eqnarray}\label{eq:eqangle2}
\sin{2 \theta} \frac{\partial \theta}{\partial \tau_2}   = \frac{X}{2} \bigl( X b_{1}-a_{1}+a_{2} -  \nonumber \\ X \left( b_{1} + b_{2} \right) \cos{\Delta} \sin{2 \theta} \bigr) \sin{\Delta} \sin{2 \theta}  \nonumber  \\
\sin{2 \theta} \frac{\partial \Delta}{\partial \tau_2}   = X \bigl[ \left( X b_{1} - a_{1} +a_{2} \right) -  \\
X \left( b_{1} \cos^{2}{\Delta}-2 b_{2} \left( 2 - \cos^{2}{\Delta} \right) \sin{2 \theta} \right) \bigr] \cos{2 \theta} \nonumber
\end{eqnarray}

These equations also may be solved in terms of non-smooth functions \cite{PILIPCHUK199643}.
We will use equations (\ref{eq:eqangle2}) for the numerical verification of the trajectories, which will be found on the phase portraits at the different levels of excitations $X$.

Before starting the study of the phase portraits, one should notice, that the excitation level $X \sim  10^{-3}$ corresponds to the amplitude of CNT oscillations $W \sim 4. 10^{-3}$, that is appropriate for using of the elastic thin shell theory.

\begin{figure}
\centering{
(a)\includegraphics[width=40mm]{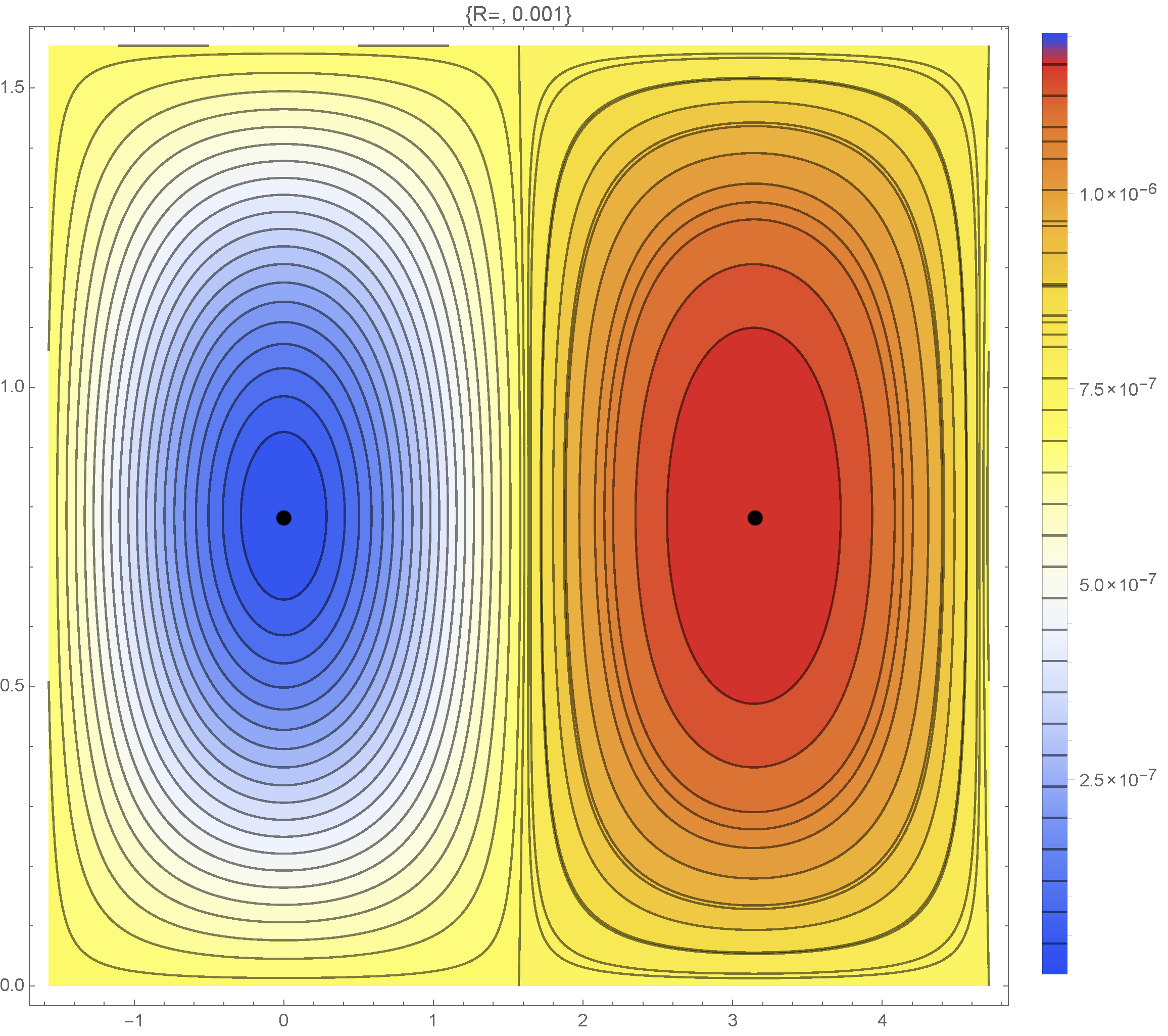} \quad (b) \includegraphics[width=40mm]{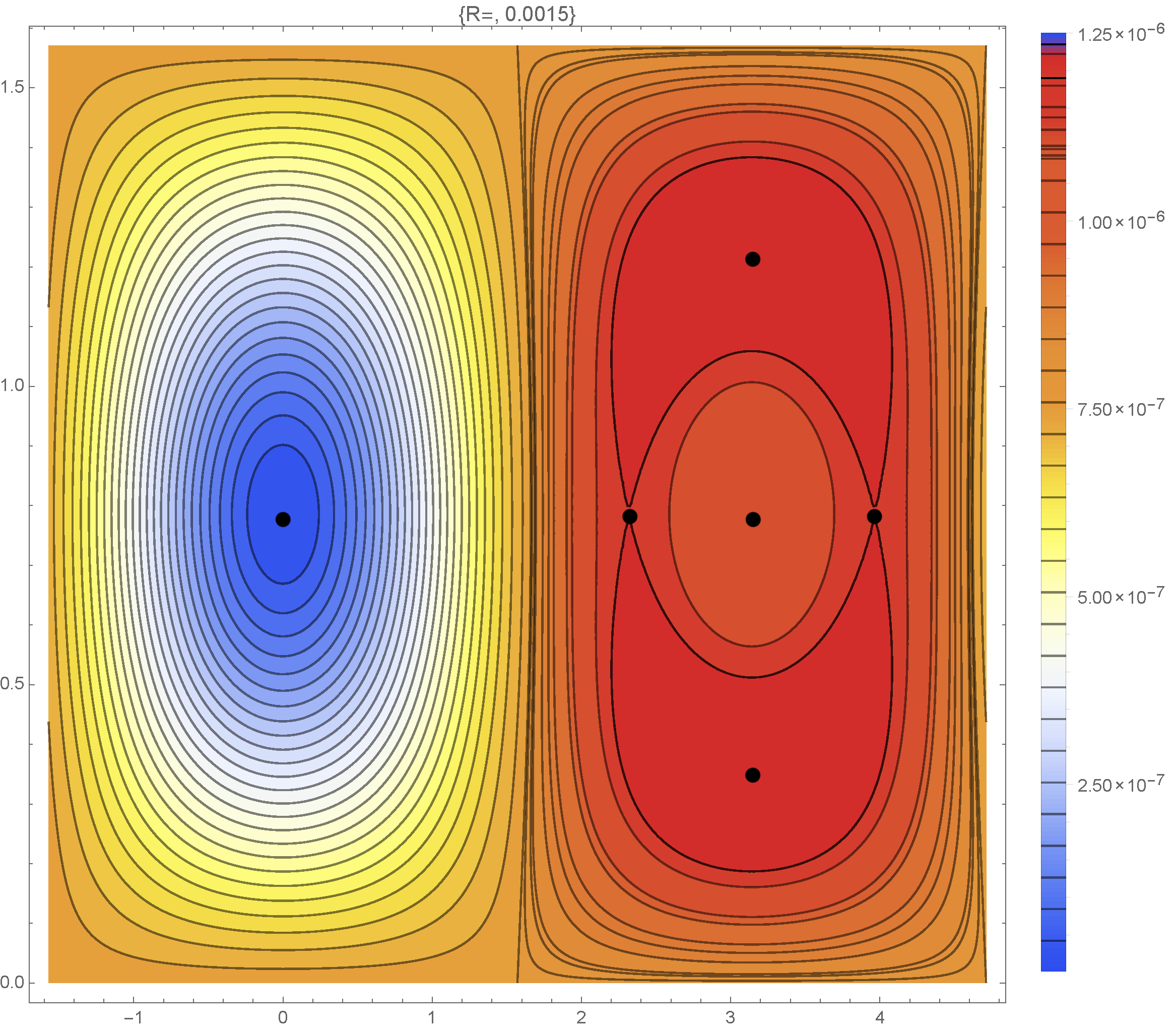} \\
 (c) \includegraphics[width=40mm]{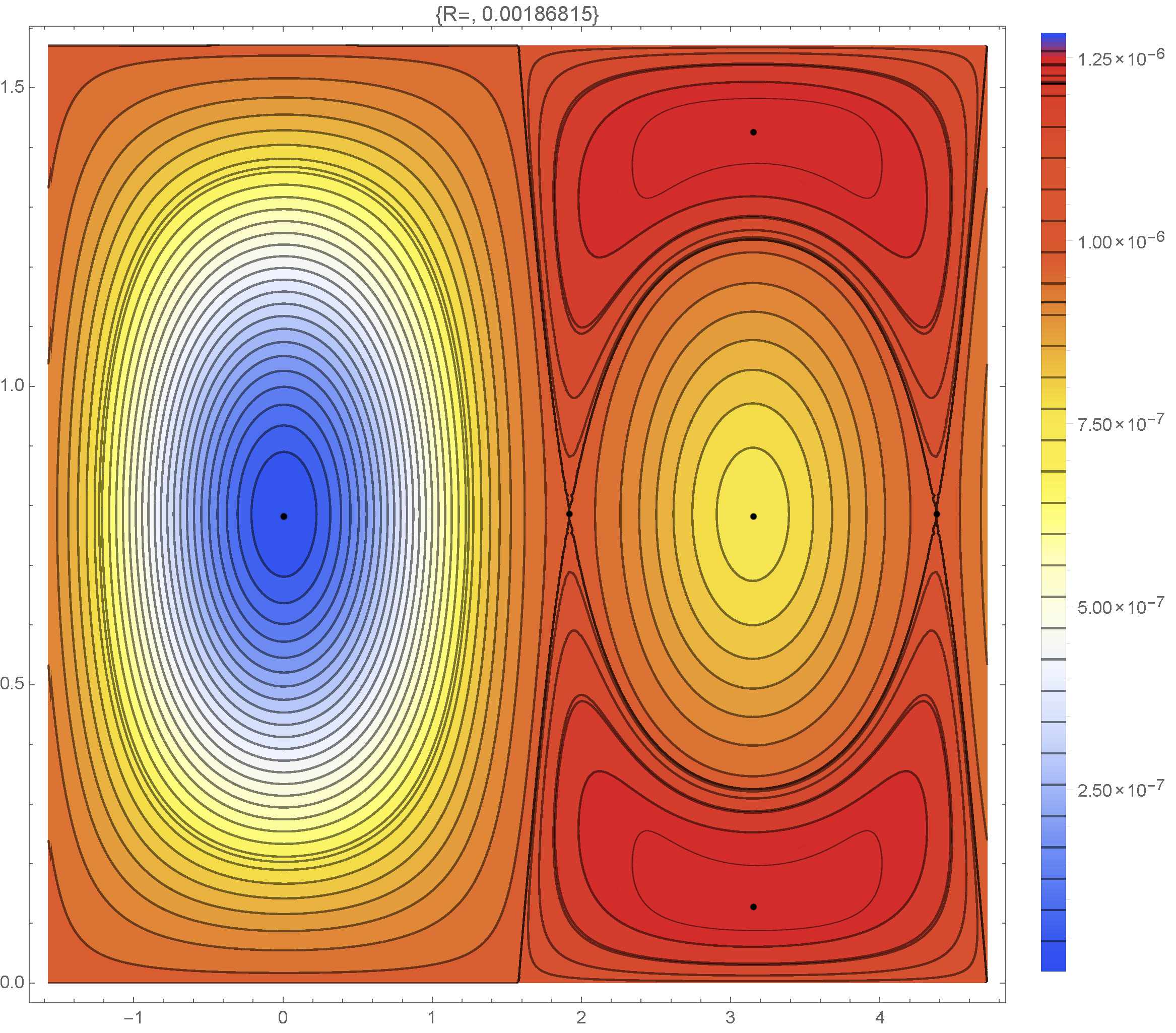} (d)\includegraphics[width=40mm]{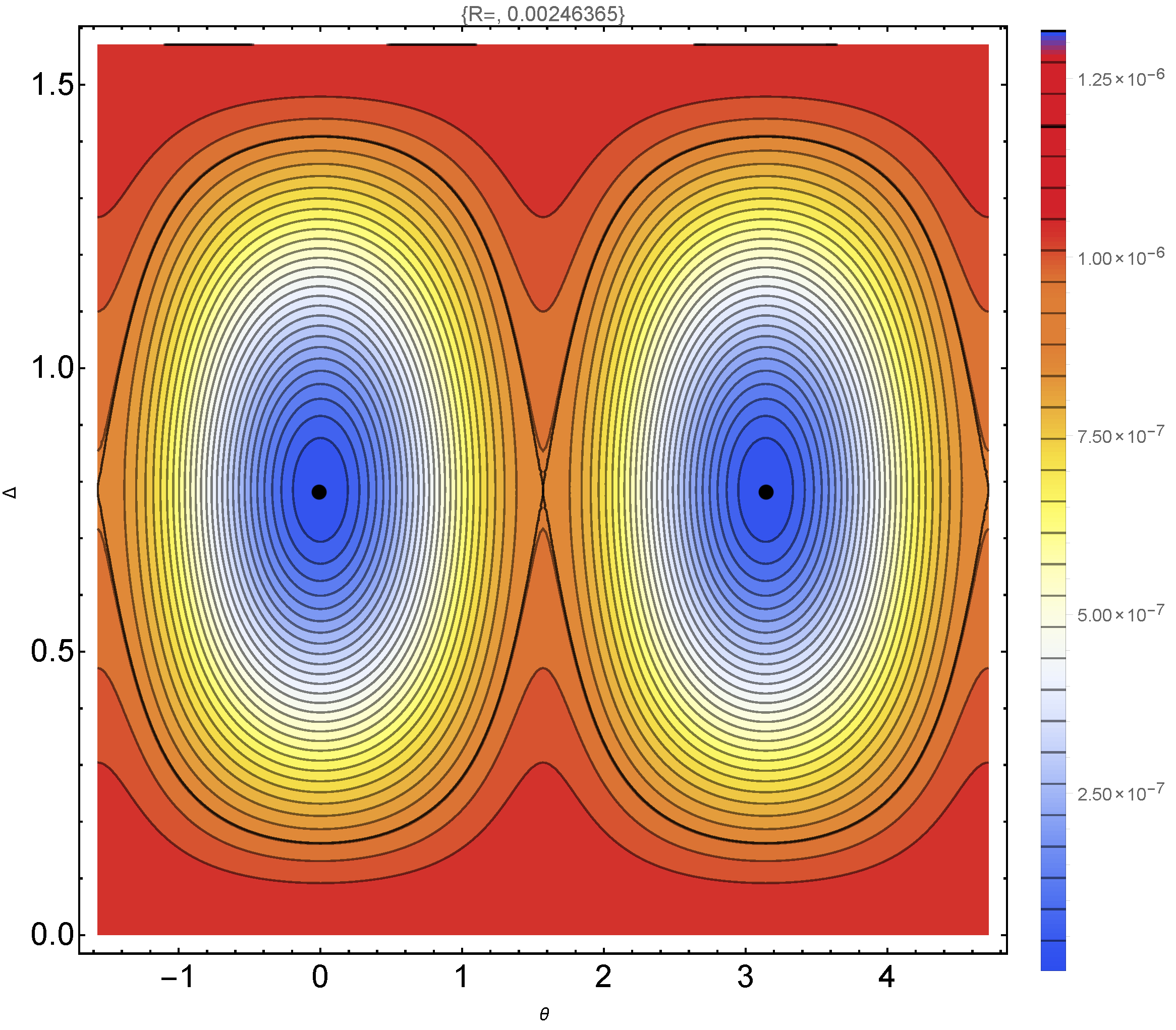}  \\
 (e)\includegraphics[width=40mm]{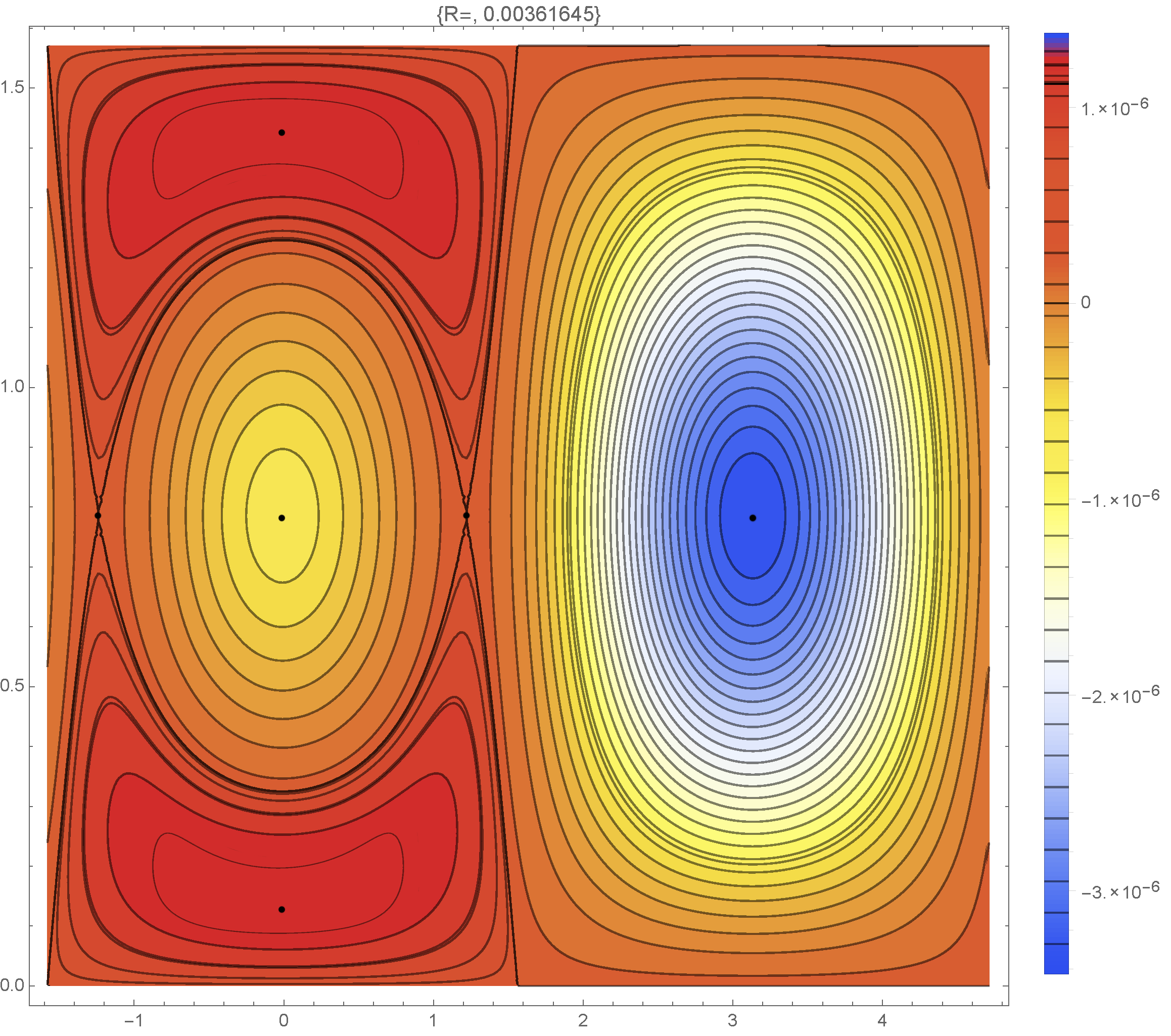} \quad (f)\includegraphics[width=40mm]{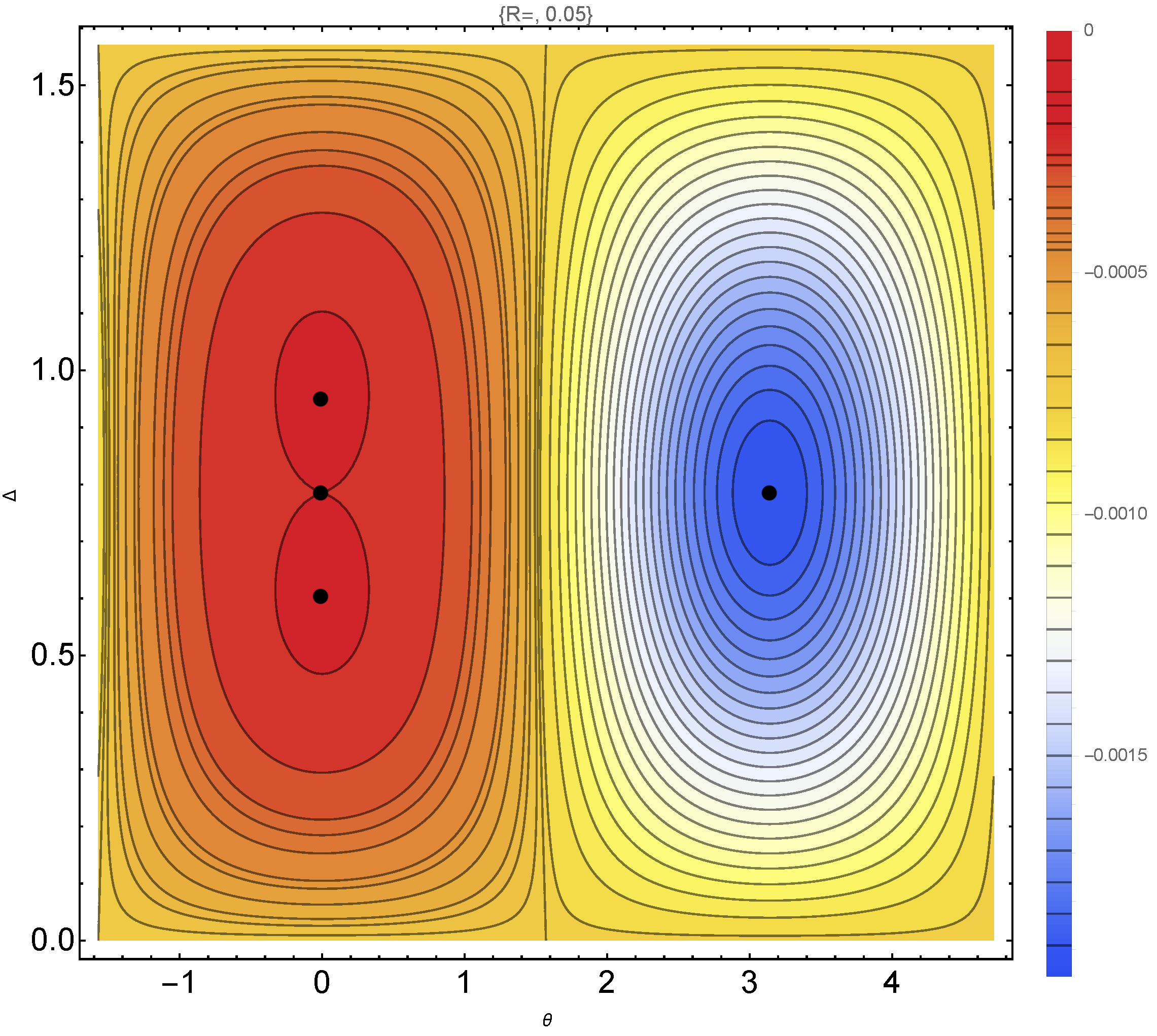}
\caption{(Color online) Phase portraits in the variables ($\Delta, \theta$) at different values of occupation number $X$: (a) $X=0.001$, (b) $X=0.0015$, (c) $X=0.00186815$, (d) $X=0.00246$, (e) $X=0.00361645$, (f) $X=0.05$.}
}
\label{fig:PP1}
\end{figure}

The analysis of the system can be performed by the phase portrait method.
Figure \ref{fig:PP1}(a) shows the phase portrait in terms of variable $(\theta, \Delta)$ for the small excitation level, which corresponds to the occupation number $X=0.001$.

The topology of the phase portrait is defined by the presence of two stationary points ($\{\Delta =0, \theta=\pi/4\}$ and $\{\Delta=\pi, \theta=\pi/4\}$), which correspond to the
normal BLOs and CFOs (eq. (\ref{eq:psistat}) ), respectively.
Any trajectories surrounding these stationary states associate with a combination of the BLOs and CFOs.
In particular, the trajectory passing through the states with $\theta=0$ and $\theta=\pi/2$ corresponds to the "domain" variables (\ref{eq:domains1}) and separates the attraction areas of the NNMs.
This trajectory is most remote from the stationary point and it is called the Limiting Phase Trajectory (LPT).
One should note also that the motion along the LPT is accompanied with the transformation of the energy distribution as it is shown in figures \ref{fig_energymap1}(c) and \ref{fig_energymap1}(d).
However, due to  expression (\ref{eq:hamiltonian2}) is the nonlinear function of the occupation number $X$, the topology of the phase portrait can be changed while the value of $X$ is varied.

\begin{figure}\centering{
\includegraphics[width=70mm]{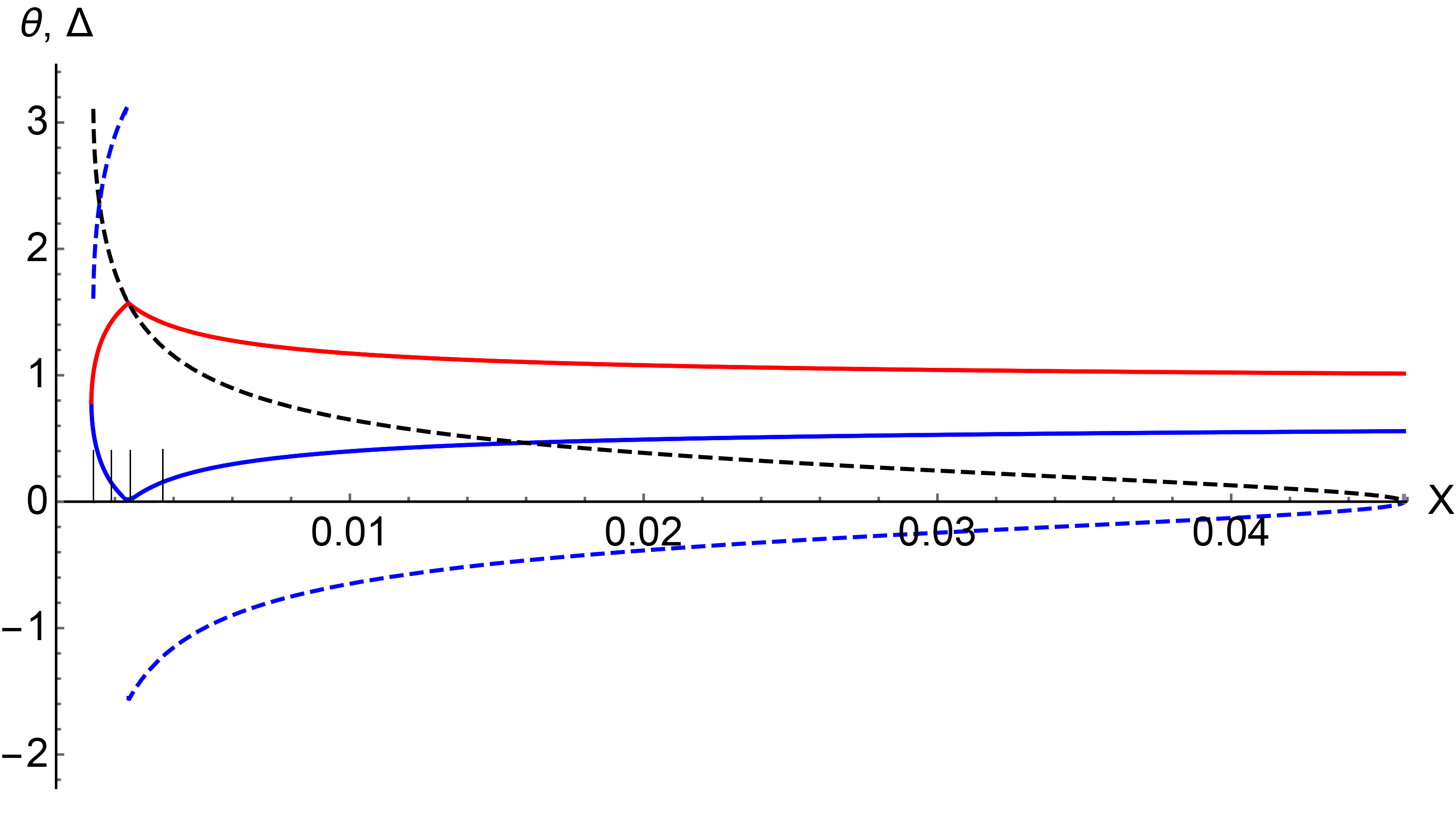}
\caption{(Color online) The stationary points' coordinates ($\theta$ and $\Delta$) vs occupation number $X$. Black and blue dashed curves show the $\Delta$-coordinates for the unstable  stationary points with $\theta=\pi/4$; solid blue and red curves show the $\theta$-coordinates for the stable stationary points with $\theta \ne \pi/4$ and $\Delta=0$ and $\pi$, respectively. Thin vertical lines represent the bifurcation values of $X$. CNT's parameters: $\alpha = 1/30,\, \beta = 0.08, \, \nu=0.19$.}
}
\label{fig:TDvsX}
\end{figure}

The analysis shows that several qualitative transformations of the phase portrait occur while the occupation number $X$ grows.
Figure \ref{fig:TDvsX} shows the values of the stationary points determining topology of the phase portrait in dependence of the occupation number $X$.

There are two stable stationary states, which correspond to the NNMs at the small values of $X$.
The first bifurcation happens when the stationary state ($\Delta = \pi, \, \theta = \pi/4$) losses its stability along the $\theta$-direction:

\begin{equation}
\frac{\partial^{2} H}{\partial \theta ^{2}}_{|\Delta=\pi,\, \theta=\pi/4} = 0 
\end{equation}

This bifurcation occurs at

\begin{equation}\label{eq:X1}
X=\frac{a_{1}-a_{2}}{2(b_{1}+b_{2})}.
\end{equation}

The value of parameter $X$ is equal to $0.001197$ at the current parameters of the CNT.
Simultaneously, two additional stable states are generated on the line $\Delta=\pi$ with $ \theta \ne \pi/4$. 

However, further growth of $X$ leads to the change of the curvature along $\Delta$-direction.

\begin{equation}
\frac{\partial ^{2} H}{\partial \Delta ^{2}}_{|\Delta=\pi,\, \theta=\pi/4} = 0
\end{equation}

In such a case the steady state $(\Delta = \pi,\, \theta=\pi/4)$ becomes stable, but two new unstable stationary point on the line $\theta= \pi/4$ arise with $\Delta \ne \pi$.
The respective value of $X$ can be estimated by the relation:

\begin{equation}\label{eq:X2}
X=\frac{a_{1}-a_{2}}{2(b_{1}-b_{2})}.
\end{equation}
 The respective value of the parameter $X$ is equal to $0.00126$.
 
Figure \ref{fig:PP1}(b) shows the phase portrait after these bifurcations ($X=0.0015$).
One can observe that four additional stationary points, which correspond to new NNMs,  appear in the quadrant ($\pi/2 \le \Delta \le 3\pi/2, \, 0 \le \theta \le \pi/2$).
Two separatrixes passing through the unstable stationary points bound the areas with the limiting variations of the amplitudes.
The stationary points with $\Delta = \pi$ correspond to the stable localization of the energy along the azimuthal angle, while the motion along the separatrix leads to fast change of the energy distribution. 
At the time the trajectories, which are close to the LPT, preserve passing between the states $\sigma_{1}$ and $\sigma_{2}$.
However, the areas which are surrounded by the separatrixes, are enlarged while the parameter $X$ grows.
The separatrixes coincide with the states ($\sigma_{1}$, $\sigma_{2}$) when their energies become to be equal to the energy at the unstable stationary points. 
It happens when the occupation number $X$ satisfies the relation

\begin{equation}\label{eq:X35}
X=\frac{\left(a_{1}-a_{2} \right) \left(b_{1} \pm \sqrt{-b_{1} (b_{1}+2 b_{2})} \right)}{\left(b_{1}+2 b_{2} \right)^{2}+4 b_{2}^{2}}
\end{equation}

Figure \ref{fig:PP1}(c) shows the phase portrait when $X$ reaches the lower value (sign $+$ in equation (\ref{eq:X35}), which is equal to $0.00187$.
At this moment the LPT coincides with the separatrix and no trajectory, which couples the $\sigma_{1}$ and $\sigma_{2}$ states, occurs.

\begin{figure}
\centering{
a) \includegraphics[width=50mm]{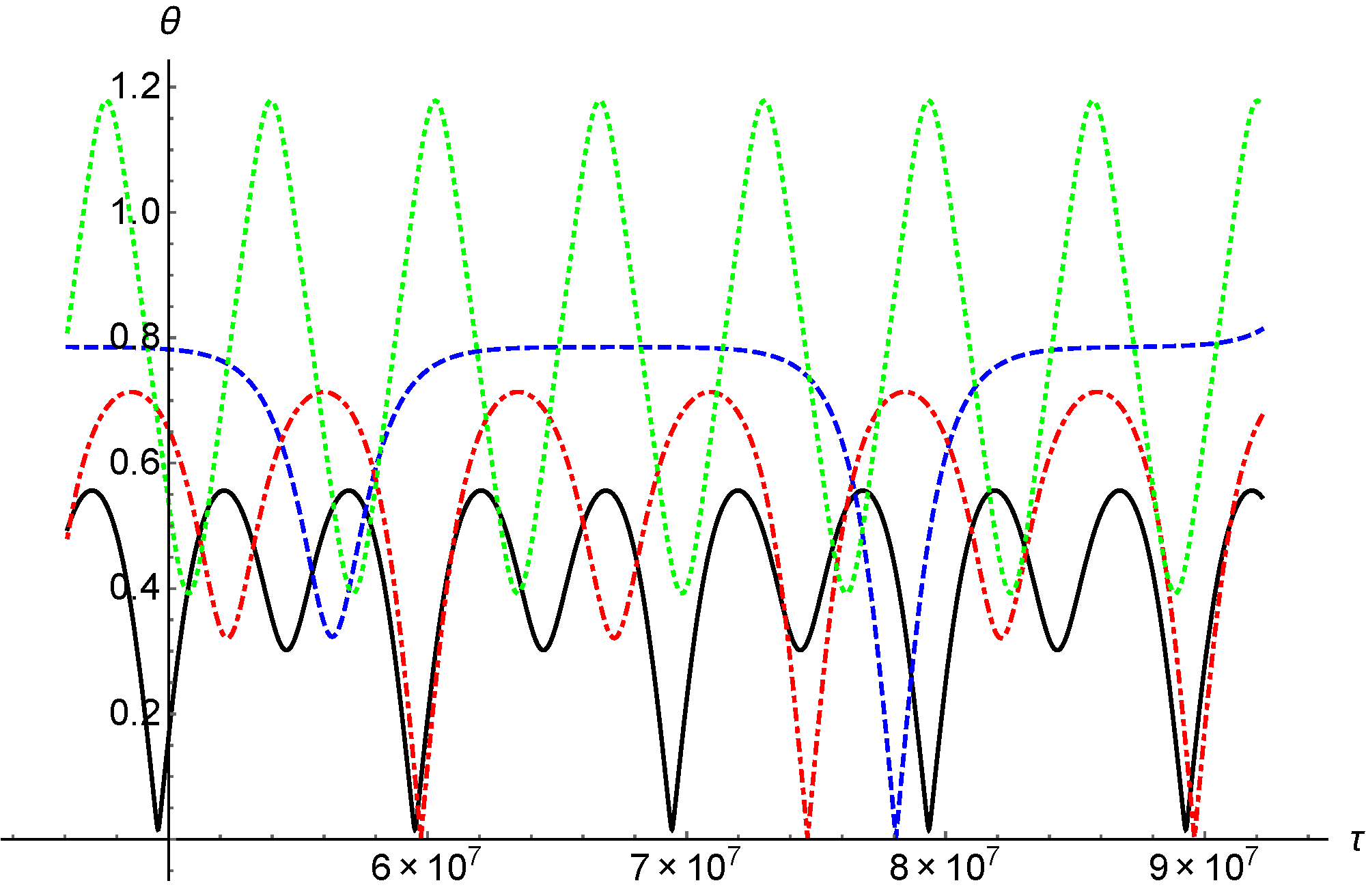} \quad b) \includegraphics[width=50mm]{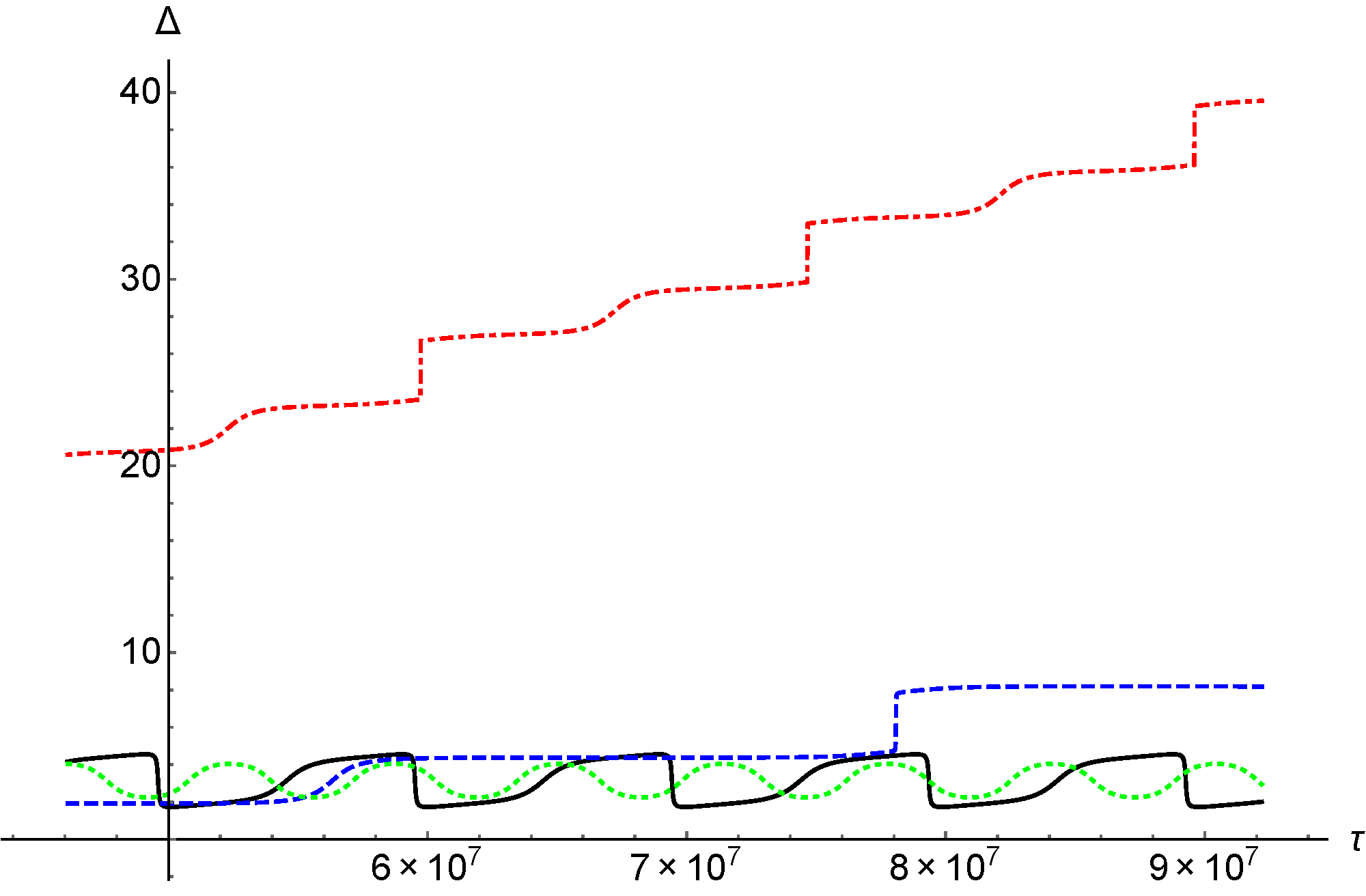}
\caption{(Color online) Time evolution of the specific trajectories $\theta(\tau_{2}),\,\Delta(\tau_{2})$ corresponding to the different initial conditions on the phase portrait. The occupation number $X=0.001868$ (the bifurcation value). Black and red dot-dashed curves show the non-stationary "localized" solutions corresponding to LPT and transit-time trajectory, respectively. Green dotted curves show some trajectory, surrounding the normal mode and blue dashed curves correspond to the trajectory passing the unstable stationary point. $\theta$ and $\Delta$ are measured in rad and the time $\tau$ in units of the oscillation period $2\pi/\omega$.}
}
\label{fig:angles000186}
\end{figure}

Two classes of the trajectories, which lead to the approximately equivalent energy distribution on the CNT surface, arise.
The first of them contains the trajectories, which surround the stable stationary points with $\Delta=\pi$ and $\theta \ne \pi/4$.
The motion along such trajectories is confined within some area, which is bounded by the LPT passing through the "domain" states $\sigma_{1}$ (or $\sigma_{2}$).
The second class is represented as a set of the transit-time trajectories, the "amplitudes" $\theta$ of which can change up to $\pi/4$, but the phase $\Delta$ grows indefinitely.
At the same time, the stationary states with $\Delta =\pi$ and $\theta \ne \pi/4$ also occur (see Fig. \ref{fig:PP1}(d)).
These states correspond to some stationary energy distribution on the CNT surface.
Figures \ref{fig:angles000186} show the time evolution of the trajectories on the Fig. \ref{fig:PP1}(c), which correspond to the different points on the phase portrait.
The black and red dashed curves describe the variables $\theta$ and $\Delta$ corresponding to the non-stationary solutions.

Next transformation happens when the "amplitude" $\theta$ of the steady states reaches the "domain" value ($\theta=0$ or $\theta=\pi/2$):

\begin{equation}
X=\frac{a_{1}-a_{2}}{b_{1}}
\end{equation}
($X=0.00246$ at the current parameters of the CNT).
At this moment the stable stationary states with $\Delta=\pi$ disappear and their analogues appear with the phase shift $\Delta=0$.

\begin{figure}
\centering{
a) \includegraphics[width=50mm]{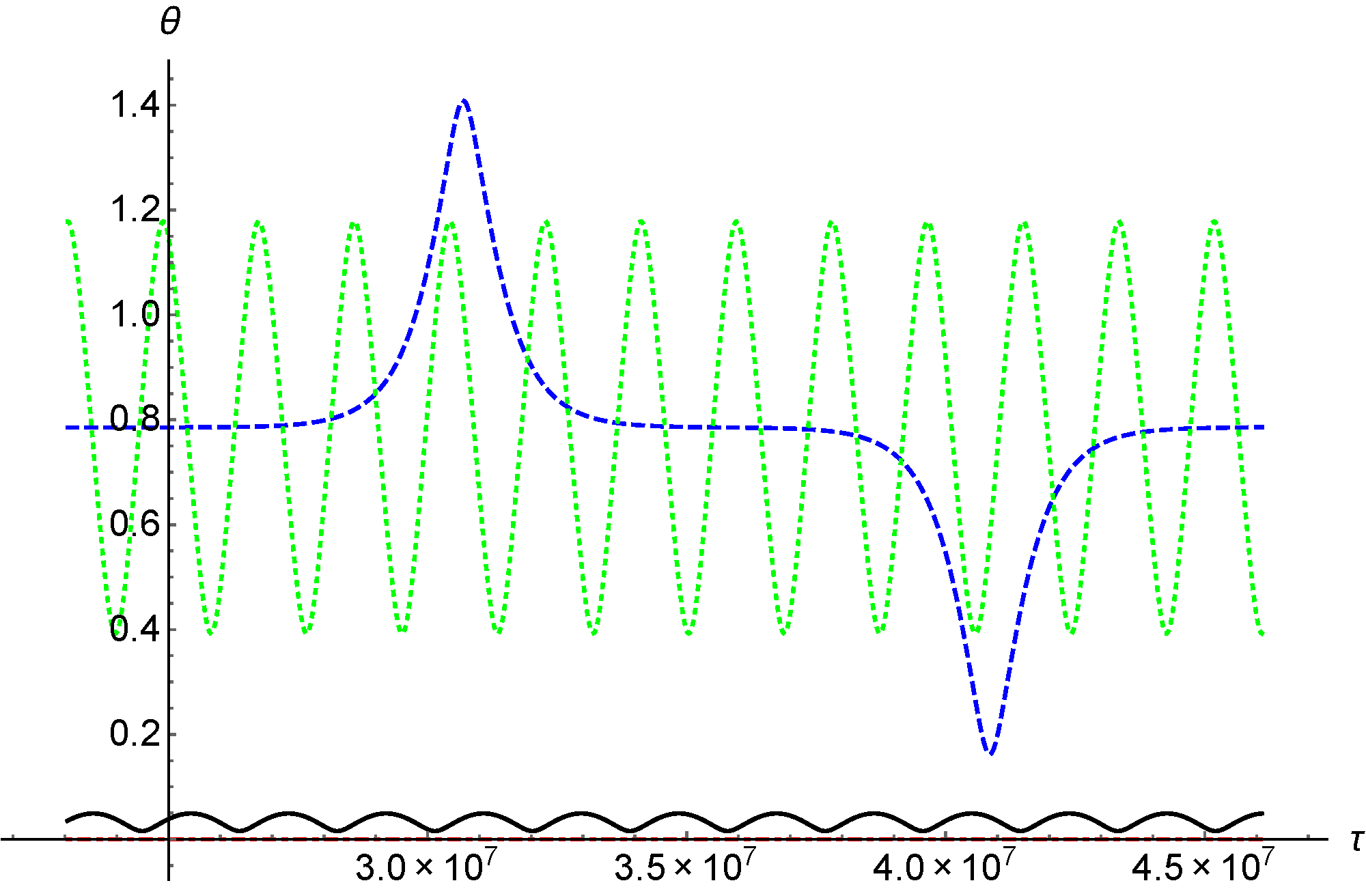} \quad b) \includegraphics[width=50mm]{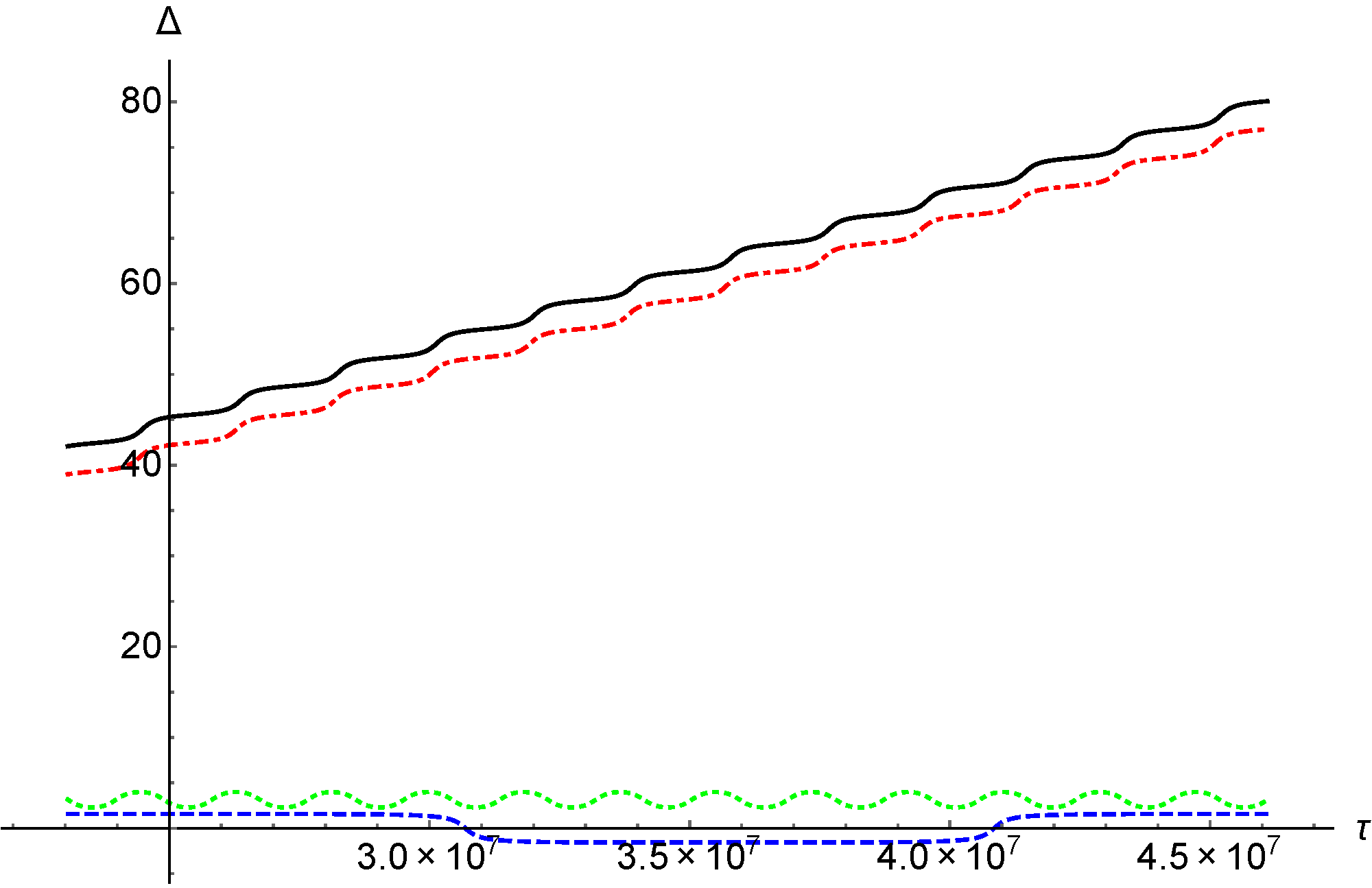}
\caption{(Color online) The same as in Fig. \ref{fig:angles000186} for the occupation number $X=0.00246$.}
}
\label{fig:angles00025}
\end{figure}

After that, while the parameter $X$ grows, the topology of the phase portrait in the quadrant $-\pi/2 \le \Delta \le \pi/2, 0 \le \theta \le \pi/2$ transforms similarly to the previous changes, but it develops in inverse direction.
The transit-time trajectories disappear at the value $X$ which is determined by equation (\ref{eq:X35}) with the sign "-" ($X=0.003617$).
After that the possibility of the full exchange between domains $\sigma_{1}$ and $\sigma_{2}$ arises again (Fig. \ref{fig:PP1}(e)).

Finally, the unstable stationary points with $\theta = \pi/4$ disappear at $X=0.0459$, but simultaneously the stationary point $(\Delta =0, \theta = \pi/4)$ becomes unstable.
New stationary states with $\Delta = 0$ and ($\theta < \pi/4, \, \theta > \pi/4$) appear, but their evolution is not interesting from our point of view (Fig. \ref{fig:PP1}(f)).

The character of the CNT oscillations is defined by the value of the occupation number $X$ and the initial conditions ($\Delta(0), \theta(0)$).

Taking these values one can integrate equations (\ref{eq:eqangle2}) numerically and then reinstate the displacement field $(u, v, w)$.

Figures \ref{fig:angles000186} and \ref{fig:angles00025} present the examples of the behaviours of the angle coordinates $\theta$ and $\Delta$ corresponding to the different trajectories on the phase portrait for two values of the parametr $X$.
Figures \ref{fig:angles000186}(a, b) correspond to the value of $X$ (\ref{eq:X35}) when the energy exchange between "domains" $\sigma_{1}$ and $\sigma_{2}$ disappears.
The non-stationary behaviour corresponding to the passing along the LPT and transit-time trajectories (black and red curves) couples with the variation of the angles $\theta, \, \Delta$ with the large period $T \sim 1.5 \, 10^{7}$.
Such a large period is explained by that these trajectories are in the vicinity of the separatrix.
One should pay the attention that the phase $\Delta$ grows indefinitely for the transit-time trajectories, while it is bounded for the LPT.
The green dotted curves on Figs. \ref{fig:angles000186} correspond to the evolution of the angles $\theta, \, \Delta$ on the trajectory, which is inside the separatrix and surrounds the circumferential normal modes $\theta=\pi/4, \, \Delta=\pi$.
Finally, the blue dashed curves show the evolution of the angles variables on the separatrix.
Therefore, their behaviour is distinguished for others essentially.
The presence of two non-identical variations of the angles corresponds to motion along the different branches of the separatrix (see Fig. \ref{fig:PP1}(d)).

The essential distinction of  Fig. \ref{fig:angles00025} from Fig. \ref{fig:angles000186} is that the variations of the angles during passing LPT as well as transit-time trajectory have the extremely small amplitudes because the value of the parameter $X=0.00246$ is approximately equal to the bifurcation value, when the stable stationary points disappears at $\Delta=\pi$.
In such a case these stationary points are extremely close to the "domain" states $\sigma_{1}$ and $\sigma_{2}$.
Therefore, the non-stationary solutions do not practically distinguished from the stationary ones.
At the same time other curves are similar to the their analogues on Fig. \ref{fig:angles000186}.

The curves on Figs. \ref{fig:angles000186} and \ref{fig:angles00025} have been calculated by the numerical integration of equations (\ref{eq:eqangle2} for two values of the parameter $X$.
Also, we have estimated the behaviour of the variables $\theta(\tau_{2})$ and $\Delta(\tau_{2})$  for other values of $X$ under various initial conditions.
In all cases the data obtained  show the excellent agreement with the structure of the phase portrait.


\section{Conclusion}

The problem of the nonlinear mode interaction is the most difficult one even in the case of well studied systems like the one-dimensional nonlinear lattices.
At the moment this problem is essential for the processes of the thermoconductivity of solids, denaturation of the DNA in biological systems, dynamics of the micro- and nano-electromechanical devices and others.
As concerns such complex objects as the thin elastic shells, in particular, modeling the CNTs, the analytic approaches to the normal modes interactions have not been formulated.
Nevertheless, the semi-inverse method used in the current work, allows to analyze the resonant mode interactions  for both the stationary and  the non-stationary processes.
We would like to emphasize that using this method make searching the normal mode profiles as well as the calculations of the frequencies of the natural oscillations very clear even for the systems, which have not any evident linearized presentation \cite{Zhang2016,Smirnov2017}.
We demonstrated high efficiency of analyzing the non-stationary dynamics without any \textit{a priori} given small parameter.
As the result, we could study the extremely complex dynamics of the CNT under condition of the resonant interaction of the NNMs, which correspond to the different branches of the spectrum.
One should note that this approach with some simplifications have been used for the study of the energy exchange and localization for the optical-type oscillations of the CNTs \cite{Smirnov2014,Smirnov2016PhysD}.

The dynamical energy redistribution and localization along the circumferential coordinates in the CNTs as well as in the macroscopic thin elastic shells may be interesting from the different points of view.
The resonant interaction of the beam-like and circumferential flexure oscillations leads to the extremely complex topology of the phase space of the system.
In particular, the nonlinear normal modes, which correspond to the energy concentration in the certain domain of the azimuthal coordinates, appear at the bounded range of the oscillation amplitudes as well as the processes of the full energy exchange between different quadrants of the shell.
Such a behaviour does not correlate with energy exchange and localization at the resonant coupling the NNMs, which correspond to the same branch of the dispersion relations.

\begin{acknowledgements}
Authors are grateful to Russia Science Foundation (grant 16-13-10302) for the financial supporting of this work.
\end{acknowledgements}

\bibliographystyle{spmpsci}      
\bibliography{Beam&CFM}   

%
%

\end{document}